\documentstyle[aps,preprint,tighten,feynmp,epsf]{revtex}
\def\be{\begin{eqnarray}}
\def\ee{\end{eqnarray}}
\def\bc{\begin{center}}
\def\ec{\end{center}}
\def\r{\rho}
\def\ro{\r_0}
\def\om{\omega}
\def\lb{\lambda}
\def\Lb{\Lambda}

\def\mpi{m_{\pi}}
\def\mn{m_N}
\def\vk{\vec{k\,}}

\def\hk{\hat{k}}
\def\hp{\hat{p}}
\def\Re{{\rm Re\,}}
\def\Im{{\rm Im\,}}
\def\prt{\partial}
\def\Sp#1{{\rm Tr}\left\{#1\right\}}

\def\gsim{\stackrel{\scriptstyle >}{\phantom{}_{\sim}}}
\def\tilde#1{\widetilde{#1}}
\begin{document}
\begin{fmffile}{diag}
%%%%%%%%%%%%%%%%%%%%%%%%%%%%%%%%%%
\title
{ Strangeness Modes in Nuclei Tested by Anti-Neutrinos }
%%%%%%%%%%%%%%%%%%%%
\author{E.E. Kolomeitsev$^a$\thanks{
e-mail: E.Kolomeitsev@gsi.de}, and D.N.
Voskresensky$^{a,b}$\thanks{ e-mail: voskre@rzri6f.gsi.de}}
\address{$^a$Gesellschaft f\"ur Schwerionenforschung, Planckstr. 1,
64291 Darmstadt, Germany\\ $^b$Moscow Engineering Physical Institute,
Kashirskoe shosse 31, 115409 Moscow, Russia}
\maketitle
%%%%%%%%%%%%%%%%%%%%%%%%%%%%%%%%
\begin{abstract}
Production of negative strangeness in reactions of inelastic
anti-neutrino scattering on a nucleus provides information on the
modification of strange degrees of freedom in nuclear matter. We
calculate cross-sections of the reaction channels
$\bar{\nu}_{e(\mu)}\rightarrow e^+(\mu^+) + K^{-}$ and
$\bar{\nu}_{e(\mu)} + p \rightarrow \Lb + e^+(\mu^+)$ and
investigate their sensitivity to the  medium effects. In
particular, we consider effects induced by the presence of a
low-energy excitation mode in the $K^-$ spectrum, associated with
the correlated $\Lambda$-particle and proton-hole states, and the
renormalization of the weak interaction in medium.  In order to
avoid double counting, various contributions to anti-neutrino
scattering are classified with the help of the optical theorem,
formulated within the non-equilibrium Green's function technique.
\end{abstract}
%\centerline{Version of \today}
\thispagestyle{empty}
\vspace*{1.5cm}


\newpage
\setcounter{page}{1}
\section{Motivation.}
The knowledge of  strange particle properties in nuclear matter is
of importance for the many interesting phenomena. For example,
hyperonization and $K^-/\bar{K}^0$ condensation in neutron
stars~\cite{kcond,kvknucl,hyper}, enhancement of $K^-$ yield in
heavy-ion collisions~\cite{hic,kvkijmp}, scattering of strange
particles on nuclei~\cite{knucl}, and level shifts in kaonic atoms
\cite{katom} are discussed recently in the literature. To gain new
information it is desirable to design  experiments, which directly
probe the  in-medium modification of strange particle properties.
In Ref.~\cite{sawyerkaon} R.F. Sawyer suggested to study  the
reaction $\bar{\nu}_{e(\mu)}
\rightarrow e^+(\mu^+) +K^{-}$, decay of an anti-neutrino in a nucleus
into a positive lepton  and an in-medium kaon. This process can
occur only if a kaon with space-like momentum can propagate in
nuclear matter and, therefore, would demonstrate that the kaon
spectrum is modified in medium compared to its vacuum form. In
Ref.~\cite{sawyerkaon}  R.F. Sawyer described the $K^-$ spectrum in
terms of a single quasiparticle  mode, $\om_{K^-}(k)$, determined
by attractive scalar and vector potentials. Consequently, for
momenta exceeding a critical value $k_c$, the spectrum is soft with
$\om_{K^-}(k)\le k$. Hence, for an anti-neutrino with the energy
$E_{\bar \nu}>k_c$ the above reaction channel opens. The value of
the critical momentum is $k_c\approx 2000\, \mbox{ MeV}-\frac43\,
\Sigma_{KN}$, where $\Sigma_{KN}$ is the kaon--nucleon
$\Sigma$-term. Using the range of $\Sigma_{KN}$  from
Ref.~\cite{kcond}, $200\,
\mbox{ MeV}<
\Sigma_{KN}< 400\, \mbox{ MeV} $, we estimate the critical momentum as
$1500\, \mbox{ MeV}<k_c<1700\, \mbox{ MeV}$. At such large momenta,
the description of the kaon--nucleon interaction in terms of
potentials becomes questionable. Furthermore, one should include
momentum dependent terms in the kaon self-energy, which were not
considered in Ref.~\cite{sawyerkaon}. A more realistic modification
of the self-energy could push the critical momentum $k_c$ to even
larger values. Below, we argue that this, indeed, happens in the
framework of a more constrained description of the $K^-$
self-energy.

As was pointed in Ref.~\cite{kvknucl} the $K^-$ spectrum shows a
second  branch related to the correlated $\Lambda(1116)$ particle
and proton hole states with the quantum numbers of $K^-$ mesons.
The typical energy of this mode is $\omega \sim m_\Lb-\mn\simeq
200$~MeV. This low-lying branch is expected to manifest itself in
neutron stars through a condensation of negative kaons with the
finite momentum (p-wave condensation) \cite{kvknucl} and in
nucleus--nucleus collisions through an enhanced population of $K^-$
modes, cf. Ref.~\cite{kvkijmp}.

On the low-lying branch in the $K^-$ spectrum the condition $\om<k$
is fulfilled at rather moderate  kaon momenta ($k_c\simeq
200$~MeV). Therefore, it is reasonable to apply the idea of R.F.
Sawyer to a new energy--momentum domain nearby this branch. In this
low-energy region $K^-$ mesons are strongly coupled with hyperons.
To constrain the $K^-$ spectral density in nuclear matter from
anti-neutrino nucleus scattering it is, therefore, necessary to
consider other channels with strangeness production in the form of
a hyperon $\bar{\nu}_{l}+N\rightarrow H +l^+$, where $N$ stands for
a nucleon and $H$ is the corresponding hyperon, $\Lb$ or $\Sigma$.
We shall show that these reaction channels give in fact a much
larger contribution to the $l^+$ cross section as compared with the
reaction channel with $K^-$ production.

In dense matter one has to consider in-medium renormalization of
the kaon and nucleon--hyperon weak currents. This renormalization
due to short-range hyperon--nucleon correlations enhances
substantially the coupling of in-medium kaons to the lepton weak
current and suppresses  the nucleon--hyperon weak current.

Due to the  strong coupling of kaons to the $\Lb$--proton-hole
states, the $K^-$ excitations are damped in matter.  Therefore, a
naive treatment of Feynman diagrams for reaction above, drawn with
the full in-medium propagators and vertices leads to double
counting. A similar problem with in-medium pion has been discussed
and resolved in Ref.~\cite{vszhetf,vossen}, with the help of the
optical theorem formalism~\cite{vossen}.  In the present work, we
apply this formalism to discriminate  various processes with
strangeness production.  Following Refs~\cite{vossen,kvaop}, we
express the strangeness production rates in terms of closed
diagrams constructed with non-equilibrium Green's functions.

Finally, we calculate cross-sections of the neutrino induced
strangeness production, utilizing the in-medium kaon spectral
density, the short range $\Lb-p$ correlations, and taking into
account the in-medium vertex renormalization. Since a neutrino can
easily pass through a nucleus and the path lengths of produced
$K^-$ mesons and $\Lb$ particles are rather small (at typical
transverse momenta under consideration), finite-size effects can be
neglected. We discuss the possibility to observe these processes in
experiment, testing, thereby, strange degrees of freedom, in
particular the kaon spectral density,  in nuclear matter.

\section{Kaon self-energy.}
Let us begin with a discussion of kaon properties in cold nuclear
matter, i.e., with the density $\rho=\rho_0=0.17$~fm$^{-3}$ and the proton
concentration $x=\rho_p/\rho=1/2$. The Green's function of the
$K^-$ meson, $D_{K^-}$, is the 
solution of the Dyson equation,
\be\label{kdyson}
\setlength{\unitlength}{1mm}
\parbox{10mm}{
\begin{fmfgraph}(10,6)
\fmfforce{(0.0w,0.5h)}{l}
\fmfforce{(1.0w,0.5h)}{r}
\fmf{boson,width=1.5thick}{l,r}
\end{fmfgraph}}
\,\,\,=\,\,\,
\parbox{10mm}{
\begin{fmfgraph}(10,6)
\fmfforce{(0.0w,0.5h)}{l}
\fmfforce{(1.0w,0.5h)}{r}
\fmf{boson}{l,r}
\end{fmfgraph}}
\,\,\,+\,\,\,
\parbox{30mm}{
\begin{fmfgraph*}(30,10)
\fmfforce{(0.0w,0.5h)}{l}
\fmfforce{(1.0w,0.5h)}{r}
\fmfpoly{empty,label=$\Pi_{K^-}$}{rd,ru,lu,ld}
\fmfforce{(0.3w,0.0h)}{ld}
\fmfforce{(0.3w,1.0h)}{lu}
\fmfforce{(0.7w,0.0h)}{rd}
\fmfforce{(0.7w,1.0h)}{ru}
\fmfforce{(0.3w,0.5h)}{kl}
\fmfforce{(0.7w,0.5h)}{kr}
\fmf{boson}{l,kl}
\fmf{boson,width=1.5thick}{kr,r}
\end{fmfgraph*}}
\,\,\, ,
\ee
 where the thin wavy line is
the Green's function of a free kaon. The thick wavy line is the
full kaon propagator in medium, which in the momentum
representation reads
$D_{K^-}(\om,k)=[\om^2-k^2-m_K^2-\Pi_{K^-}(\om,k,\r,x)+i0]^{-1}\,$.
The frequency and the momentum of a kaon are denoted by $\om$ and
$k$, respectively. The notation $m_K$ stands for the free kaon
mass, and $\Pi_{K^-}$ is the $K^-$ self-energy. The latter contains
several pieces related to the most important processes,
$\Pi_{K^-}=\Pi_S+\Pi_P+\Pi_{\rm res}$. The first term, $\Pi_S$, is
the s-wave part of the $K^{-}$ self-energy, generated by the s-wave
kaon--nucleon scattering essential nearby the $KN$
threshold~\cite{kcond,kvknucl,knswave}. The difference between
various approaches with respect to the s-wave $KN$ interaction is
reflected mainly in the description of the energy--momentum region
nearby the kaon branch of the spectrum. However, for
parameterizations \cite{knswave} considered, at present, as rather
realistic, the kaon branch lies above the line $\om=k$ up to very
high kaon momenta $k_c\sim 2000$~MeV. Therefore, we avoid a
detailed discussion of uncertainties in $\Pi_S$ term and, following
Ref.~\cite{kcond}, utilize it in a much more simple form
$\Pi_S=-d\,m_K^2\,\r/\ro-
\alpha \,m_K\,\om\,\r/\ro$, with parameters
$d\approx 0.18$ and $\alpha\approx 0.23$ taken from
Refs.~\cite{kvknucl,kvkijmp}. Further we intend to focus on another
term of the kaon self-energy, which is responsible for the spectral
density of the $K^-$ states below line $\om=k$.

The p-wave part of the $K^-$ polarization operator is mainly
determined by the contributions from the $\Lambda
(1116)$--proton-hole states and the $\Sigma(1193)$--nucleon-hole
intermediate states $\Pi_P=\Pi_\Lb+\Pi_\Sigma$. In
Ref.~\cite{kvknucl} it is argued that due to smallness of the
kaon--nucleon--$\Sigma$ coupling constant compared to  the
kaon--nucleon--$\Lb$ coupling constant
($C_{KN\Sigma}/C_{KN\Lb}\simeq 0.2$) the contribution of $\Sigma$
particles to the polarization operator is small. Therefore, we do
not consider small contributions of $\Sigma$ hyperons, which
structure is quite analogous to that of the $\Lb$ hyperon, and drop
the term $\Pi_\Sigma$ in the polarization operator.

The main contribution, $\Pi_\Lb$, is depicted by the loop diagram

$ $

\begin{equation}\label{eq:loopint}
\Pi_{\Lb} =\parbox{25mm}{
%\begin{fmffile}{loop}
\setlength{\unitlength}{1mm}
\begin{fmfgraph*}(25,15)
\fmfleft{l}
\fmfright{r}
\fmfforce{(0.0w,0.5h)}{l}
\fmfforce{(1.0w,0.5h)}{r}
\fmfforce{(0.25w,0.5h)}{ol}
\fmfforce{(0.75w,0.5h)}{or}
\fmf{boson}{l,ol}
\fmf{boson}{or,r}
\fmf{fermion,left=.5,tension=.5,label=$p$}{or,ol}
\fmf{heavy,left=.5,tension=.5,label=$\Lambda$}{ol,or}
\fmfdot{ol}
\fmfv{d.sh=c,d.f=full,d.si=4thick}{or}
\end{fmfgraph*}
%\end{fmffile}
}
= -i \, C_{KN\Lb} \,
\int\frac{d^4p}{(2\,\pi)^4}\,\tilde{C}_{KN\Lb}\,
\Sp{\hk\, \gamma_5 \, \hat{G}_{\Lb}(p+k)\, \hk\, \gamma_5\,
\hat{G}_N(p) }
\,,\end{equation}

$ $

\noindent
where $\hat{G}_{\alpha}(p)=(\hp+m_\alpha^*)\,G_\alpha(p)
=(\hp+m_{\alpha}^*)\,\left\{[p^2-m_{\alpha}^{*2}+i\,0]^{-1}
+ 2\,\pi\, i \, n_{\alpha} (p)\, \delta(p^2-m_{\alpha}^{*2})
\right\}$ is the Green's function of a given baryon, $\alpha=p,\Lb$;
$m_\alpha^*$ is the in-medium mass of the baryon $\alpha$ taken
according to Ref.~\cite{hyper},
 and
$n_{\alpha}$ stands for the Fermi occupation factor of protons. The
bare coupling constant is $C_{KN\Lb}\simeq
-1\,\mpi^{-1}$.
Here and below, evaluating integrals with the Green's functions we
drop the divergent medium-independent part, which is assumed to be
contained in the physical values of particle masses and coupling
constants.
 The fat blob in the diagram corresponds to the
full $KN\Lb$ vertex $\tilde{C}_{KN\Lb}$, which takes into account
baryon--baryon correlations. It is depicted by the following
diagrams

$ $

\begin{equation}\label{correl}
\tilde{C}_{KN\Lb}=
\parbox{11mm}{
%\begin{fmffile}{cor1}
\setlength{\unitlength}{1mm}
\begin{fmfgraph*}(11,15)
%\fmfpen{thick}
\fmfleftn{l}{2}
\fmfright{r}
\fmf{fermion}{o,l1}
\fmf{heavy}{l2,o}
\fmf{boson}{o,r}
\fmfv{d.sh=c,d.f=full,d.si=4thick}{o}
\fmflabel{$\Lambda$}{l2}
\fmflabel{$p$}{l1}
\end{fmfgraph*}
%\end{fmffile}
}
\,\,=\,\,
\parbox{11mm}{
%\begin{fmffile}{cor2}
\setlength{\unitlength}{1mm}
\begin{fmfgraph*}(11,15)
%\fmfpen{thick}
\fmfleftn{l}{2}
\fmfright{r}
\fmf{fermion}{o,l1}
\fmf{heavy}{l2,o}
\fmf{boson}{o,r}
\fmfdot{o}
\fmflabel{$\Lambda$}{l2}
\fmflabel{$p$}{l1}
\end{fmfgraph*}
%\end{fmffile}
}
\,\, +\,\,
\parbox{11mm}{
%\begin{fmffile}{cor3}
\setlength{\unitlength}{1mm}
\begin{fmfgraph*}(30,15)
%\fmfpen{thick}
\fmfleftn{l}{2}
\fmfright{r}
\fmf{fermion}{ol,l1}
\fmf{heavy}{l2,ol}
\fmf{fermion,left=.5,tension=.5,label=$p$}{or,ol}
\fmf{heavy,left=.5,tension=.5,label=$\Lambda$}{ol,or}
\fmf{boson}{or,r}
\fmfv{d.sh=s,d.f=shaded,d.si=4thick}{ol}
\fmfv{d.sh=c,d.f=full,d.si=4thick}{or}
\fmflabel{$\Lambda$}{l2}
\fmflabel{$p$}{l1}
\end{fmfgraph*}
%\end{fmffile}
}\qquad\qquad \qquad .
\\ \nonumber \\ \nonumber \\ \nonumber
\end{equation}

$ $ %\vspace*{3mm}

\noindent
The shaded square represents
 the short-range $\Lambda$--proton interaction that can be written in
the non-relativistic Landau--Migdal parameterization as

$ $

\begin{equation}\label{tln}
\parbox{20mm}{
%\begin{fmffile}{tln}
\setlength{\unitlength}{1mm}
\begin{fmfgraph*}(20,10)
%\fmfpen{thick}
\fmfleftn{l}{2}
\fmfrightn{r}{2}
\fmf{fermion}{r1,o,l1}
\fmf{heavy}{l2,o,r2}
\fmfv{d.sh=s,d.f=shaded,d.si=4thick}{o}
\fmflabel{$p$}{l1}
\fmflabel{$\Lambda$}{l2}
\fmflabel{$p$}{r1}
\fmflabel{$\Lambda$}{r2}
\end{fmfgraph*}
%\end{fmffile}
}
=
T_{\Lambda p}^{\rm loc}= C_{KN\Lb}^2
\left[ f_{\Lb}+f'_{\Lambda}\, (\vec{\sigma}_{\Lb}\,
\vec{\sigma\,}_p)\right]\, ,
%\\ \nonumber \\ \nonumber \\ \nonumber
\end{equation}

$ $

\noindent
where $\sigma_{\Lb}$ and $\sigma_p$ are the Pauli spin matrices of
a $\Lambda$ particle and a proton, respectively, $f_\Lb$ and
$f'_\Lb$ are the Landau--Migdal parameters of the $\Lb-p$
interaction. This interaction is irreducible  with respect to the
hyperon--nucleon-hole intermediate states.
 The formal solution of
Eq.~(\ref{correl}) with the interaction (\ref{tln}) is given by
\begin{equation}\label{gamma}
\tilde{C}_{KN\Lb}=\gamma(f'_\Lb)\, C_{KN\Lb}=
[1-f'_\Lb\, C_{KN\Lb}^2\, A_{p\Lb}(\om,k)]^{-1}\, C_{KN\Lb}
\end{equation}
with the loop integral $A_{p\Lb}(\om,k)=-i\,8\,\mn^{*2}\int
\frac{d^4p}{(2\,\pi)^4}
\, {G}_{\Lb}(p+k) \, {G}_N(p)$. We see that only the spin parameter
$f'_\Lb$ enters expression (\ref{gamma}). The empirical value of
$f'_{\Lb}$  is not known. It could be, in principle, extracted from
the data on multi-strange hypernuclei. For our purpose in this note
we are contented with a rough estimation of this parameter.
Following Ref.~\cite{picores}, we suggest that the hyperon--nucleon
interaction is determined mainly by the kaon and $K^*$ exchanges
corrected by the short-range $NN$ correlations
\begin{equation}\label{fprime}
f'_\Lb \simeq \frac13\,
\frac{m_0^2}{m_K^2+m_0^2} +\frac23 \,
 \frac{C_{K^*N \Lb}^2}{ C_{K N\Lb}^2} \,
 \frac{m_0^2}{m_{K^*}^{2}+m_0^2}
\,. \end{equation}
Here $m_{K^*}$ denotes the mass of the heavy strange vector meson
and $m_0$ is related to the inverse core radius of nucleon--nucleon
interaction, i.e., $m_0\simeq m_{\om}$, being the mass of the $\om$
meson. Utilizing the coupling constant of $K^*N\Lb$ interactions,
taken from the J\"ulich model of hyperon--nucleon interaction via
the meson exchange, $C_{K^*N\Lb}\simeq 1.74/\mpi$, we estimate
$f'_\Lb\approx 1.1$. The calculations below will be done for two
values of this parameter $f'_\Lb=0$ and $f'_\Lb=1.1\,$.

The last term of the self-energy, $\Pi_{\rm res}$, includes
residual interaction, which cannot be constrained from experiments
with on-shell kaons. Therefore, its value and structure is rather
ambiguous. In Refs~\cite{yabu,kvknucl} the residual off-shell
interaction is suggested to be reconstructed with the help of
low-energy theorems. These constrains, following from the current
algebra and the PCAC hypothesis, can be safely applied in our
present consideration, since the neutrino induced reaction probe
directly the axial current correlator, for which the low-energy
theorems have actually been formulated. Following
Ref.~\cite{kvknucl}, we cast the term $\Pi_{\rm res}$ as  $\Pi_{\rm
res}=\lb\,(m_K^2-\om^2+k^2)\,\r/\ro$ with the parameter
$\lb=2\,d\,$. Without regard to the low-energy theorems one would
put $\Pi_{\rm res}=0$.

Please note that the particular details of the kaon--nucleon
interaction nearby the mass-shell  (that can be found elsewhere
\cite{kcond,kvknucl,knswave}) do not affect qualitatively the
description of kaon behavior  at somewhat lower energies, which
determine our discussion below. The crucial point for us is that a
kaon couples to $\Lambda$-particle--hole states and propagates,
thereby, in a frequency--momentum region not accessible for vacuum
kaons. This coupling enforces  $K^-$ and $\Lb$ degrees  of freedom
to be treated consistently.
 We are rather going to discuss
experimental consequences of a kaon modification, using the above
formulated kaon self-energy for illustration and the short-range
$\Lb p$ interaction (\ref{tln}).

\section{$K^-$ spectral density.}
\begin{figure}[h]
\begin{minipage}[b]{0.45\textwidth}
\epsfxsize=6.5cm
\bc\mbox{\epsfbox{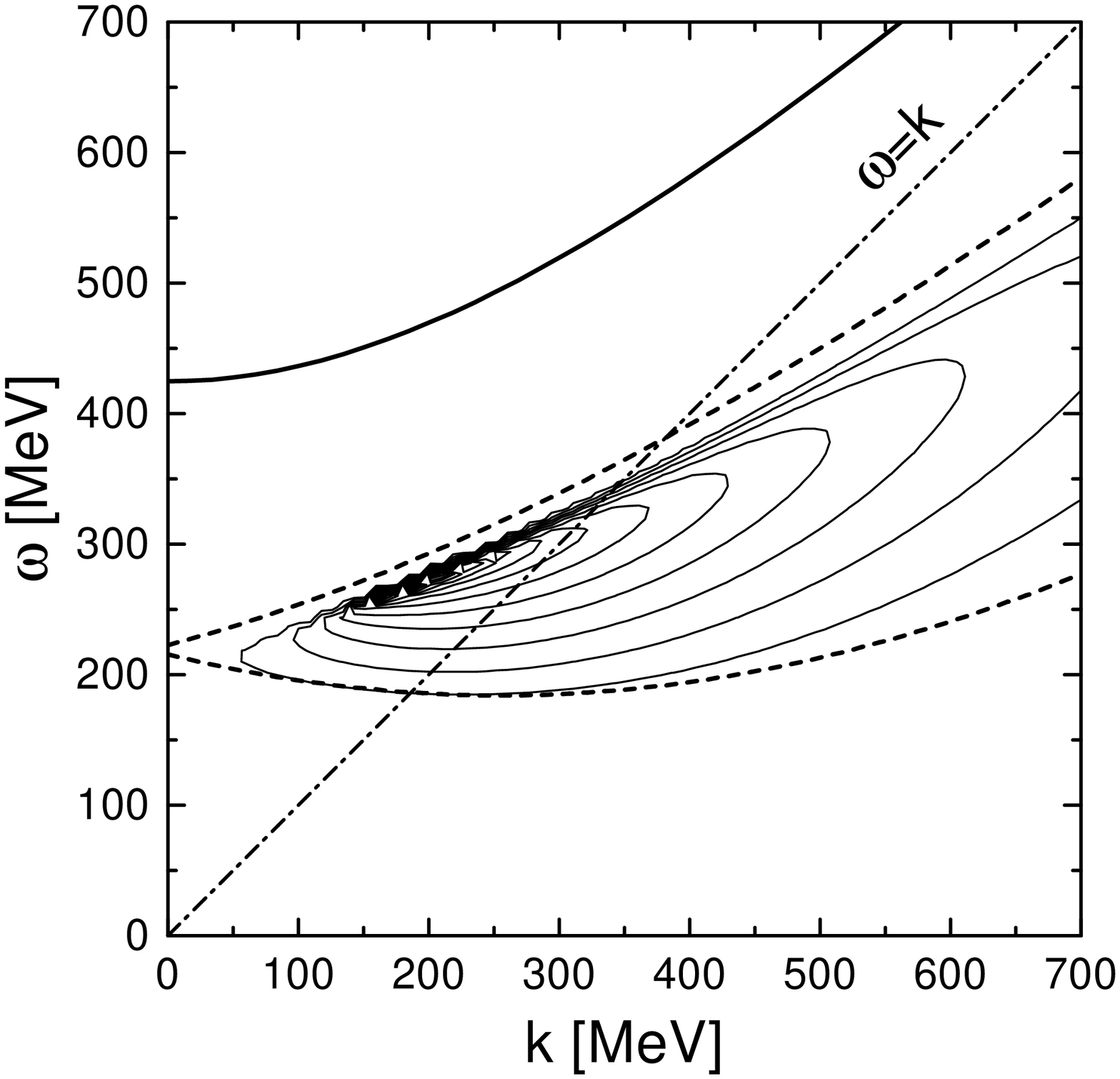}}\ec
  \end{minipage}
%  \hfill
  \begin{minipage}[b]{0.45\textwidth}
\epsfxsize=6.cm
\bc\mbox{\epsfbox{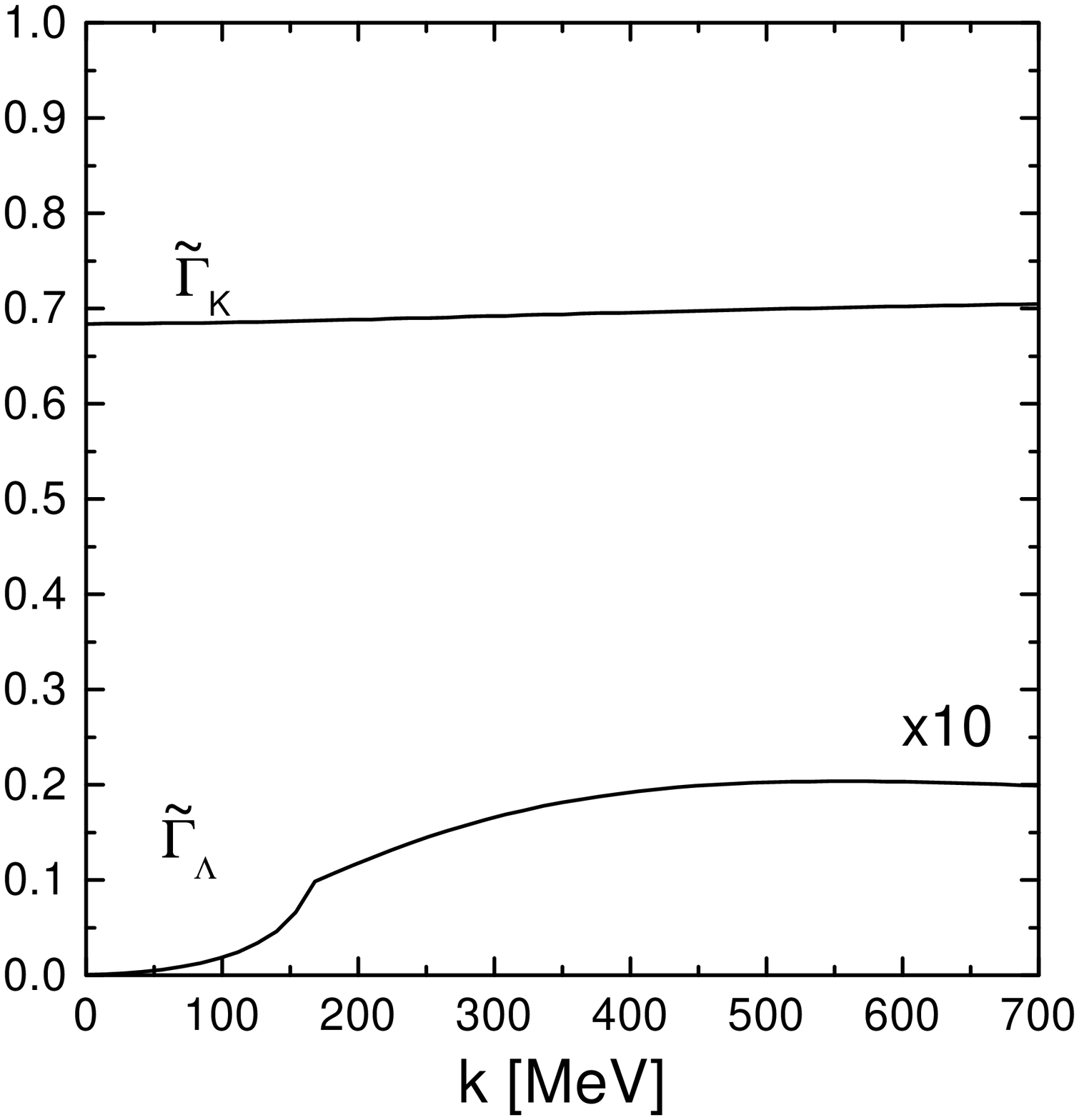}}\ec
\end{minipage}
\caption{Spectral density of $K^-$ excitations in isospin symmetrical
  nuclear matter at $\r=\ro$ (left panel) and the occupation factors of
in-medium kaons (right panel). The line $\tilde{\Gamma}_K$
corresponds to the upper kaon branch. The line $\tilde\Gamma_\Lb$
is related to the integral over the region of the
$\Lb$--proton-hole continuum between dashed lines on the left
plane.}
  \label{fig:specdens}
\end{figure}

The spectral density of kaon excitations is defined as
$A_{K^-}(\om,k)=-2\Im D^R_{K^-}(\om,k)$, where $D^R_{K^-}$ is the
retarded Green's function of the $K^-$ meson. The left panel in
Fig.~\ref{fig:specdens} shows the contour plot of $A_{K^-}(\om,k)$
calculated for $\r=\ro$ and $x=1/2$.  The upper solid line
corresponds to the quasiparticle kaon branch. In the framework of
our simplified treatment of the s-wave $K^-N$ interaction  the
spectral density is a $\delta$-function,
$A_{K^-}(\om,k)=\bigl[\tilde{\Gamma}_{K^-}(k)/2\om_{K^-}(k)\bigr]
\,\delta(\om-\om_{K^-}(k))$ for $\om$ nearby $\om_{K^-}(k)$, where
$\om_{K^-}(k)$  is the solution of the dispersion equation
$\Re[D_{K^-}^{R}(\om,k)]^{-1}=0$, which determines the kaon branch
of the spectrum. The factor
$\tilde{\Gamma}_{K^-}(k)=2\om_{K^-}(k)\, [
\prt \Re D^R_{K^-} (\om_{K^-}(k),k) / \prt\om]^{-1}$ measures how
strongly the kaon branch is populated by the  in-medium $K^-$
mesons, cf. Ref.~\cite{kvknucl}. It indicates the relative weight
of this branch in the full spectral density. The dot-dashed line on
the left panel in Fig.~\ref{fig:specdens} is the upper border of
the region $\omega <k$, where the processes under consideration may
occur. We see that the upper quasiparticle kaon branch does not
cross this border for momenta $k< k_c$, where the critical momentum
$k_c=m_K (1-d+\lb)/\alpha \approx 2570$~MeV (for $\lb=2\,d$ that
corresponds to the more constrained description) and $k_c\approx
1570$~MeV (for $\lb=0$, i.e., when one ignores constrains of the
low-energy theorems).

Below the kaon branch in Fig.~\ref{fig:specdens} are shown the
contour lines of the kaon spectral density in the
$\Lambda$--proton-hole continuum. The latter is bordered by the
dashed lines. Within the continuum the imaginary part of the kaon
propagator is non-zero, $\Im\Pi_\Lb\neq 0$, that corresponds to
processes $K^-\leftrightarrow \Lb + p^{-1}$ ($p^{-1}$ means the
proton hole). The relative strength of this region in the spectral
density is characterized by the quantity
\be\label{gammalb}
\tilde{\Gamma}_{\Lb}(k)=\intop_{\om_{p\,\Lb}^-(k)}^{\om_{p\,\Lb}^+(k)}
\frac{d\om}{2\pi}
2\, \om A_{K^-}(\om,k)\,,
\ee
where $\om_{p\,\Lb}^\pm(k)$ is the upper ($+$) and
lower ($-$) borders of $\Lb\,p^{-1}$ continuum,
\be \nonumber
\om_{p\,\Lb}^+(k)
&=&\Biggl\{
\begin{array}{lr}
\sqrt{(m_\Lb^*-\mn^*)^2+k^2} &, \quad k<p_{p\,{\rm F}}\, (m_\Lb^*-\mn^*)/\mn^*
\\
\sqrt{m_{\Lb}^{*\,2}+(k+ p_{p,{\rm
F}})^2}- \sqrt{\mn^{*\,2}+p_{p,{\rm F}}^2} &, \quad k>
p_{p\,{\rm F}}\, (m_\Lb^*-\mn^*)/\mn^* \end{array}
\,,
\\ \nonumber
\om_{p\,\Lb}^-(k)
&=&\sqrt{m_{\Lb}^{*\,2}+(k- p_{p,{\rm
F}})^2}- \sqrt{\mn^{*\,2}+p_{p,{\rm F}}^2}
\,,\ee
 $p_{p,{\rm F}}$
stands for the Fermi momentum of the protons. The values of the
occupation factors $\tilde{\Gamma}_{K^-}(k)$ and
$\tilde{\Gamma}_{\Lb}(k)$ are shown in the right panel of
Fig.~\ref{fig:specdens}. We observe that the main weight is carried
by the kaon branch, whereas the lower $\Lb\,p^{-1}$ continuum is
populated by $K^-$ only on a percentage level.

Please notice that in the quasiparticle limit, $\mbox{Im} \Pi
\rightarrow 0$, our kaon spectrum has only one kaon branch and the
dispersion equation $\Re[D_{K^-}^{R}(\om,k)]^{-1}=0$ has no
low-energy solution. Only in the resonance approximation for the
real part of the particle--hole loop such a solution exists,
cf.~\cite{kvknucl}. In this case one can approximately treat the
low-energy region as a quasiparticle branch. Besides, with taking
into account the complicated threshold dynamics in $K^-N$
scattering nearby threshold, leading to the dynamical
$\Lambda^*(1405)$ resonance, the upper quasiparticle kaon branch
becomes also a region of the spectral density. Hence, in reality,
we discuss experimental manifestation of the regions on the
frequency--momentum plane, where the spectral function does not
vanish, rather than manifestation of quasiparticle branches. For
brevity's sake we continue to speak about "the branches", bearing
in mind  regions of particle--hole continua populated by mesonic
excitations.

The description of the kaon--nucleon interaction formulated above
(cf. Ref.~\cite{kvknucl}) is quite analogous to that well known
from pion--nucleon physics, cf. \cite{migdal}. Except for a spin
and an isospin, hyperons play the same role in kaon physics, as
$\Delta$ isobars do in pion physics. The crucial difference,
however, is that the corresponding $\Delta$ branch in the pion
spectrum lies above the pion branch, whereas the $\Sigma$ and
$\Lambda$ branches of the $K^-$ spectrum lie below the kaon branch.
The difference in the description  of the s-wave interactions in
the pion and the kaon cases arises mainly due to distinct values of
the corresponding $\Sigma$-terms. The pion--nucleon $\Sigma$-term
is much smaller than the kaon--nucleon one. Due to this, pion
condensation may occur at sufficiently high density $\rho\gsim
2\,\rho_0$ due to attractive p-wave interaction, as has been
suggested by A.B. Migdal,  R.F. Sawyer, and D.J. Scalapino, cf.
Ref.~\cite{picond}. On the contrary, kaon condensation may occur as
due to the s-wave attraction \cite{kcond} as due to the p-wave one
\cite{kvknucl} (the latter possibility is analogous to that in the
pion case). The choice between these possibilities depends on the
interplay between strengths of not well known s- and p-wave
interactions in dense nuclear  matter.

The neutrino reactions discussed in this paper might give an extra
important information on strange particle interaction in nuclear
matter. It could yield additional constrains on the
$K^-$--$\Lb$--nucleon interaction.

\section{Weak interaction in medium}
The kaon decay processes in vacuum are determined by the current
$J_K^{\mu}= i\,\sqrt{2}\, f_{K}\, k^{\mu} $, where  $f_K$ stands
for the kaon decay constant. In the nuclear medium this current is
modified due to strong interaction, $ \tilde{J}_K^{\mu}= i\,
\sqrt{2}\, f_{K}\, \Gamma^{\mu}(k) $, where the vertex function
 $\Gamma(k)$ is mainly determined by the following diagrams

$ $

\be\label{diagk}
\parbox{11mm}{
%\begin{fmffile}{kcor1}
\setlength{\unitlength}{1mm}
\begin{fmfgraph*}(11,15)
\fmfrightn{l}{2}
\fmfleft{r}
\fmf{scalar}{o,l1}
\fmf{scalar}{l2,o}
\fmf{boson}{o,r}
\fmfv{d.sh=di,d.f=full,d.si=4thick}{o}
\end{fmfgraph*}
}
\,\,=\,\,
\parbox{11mm}{
\setlength{\unitlength}{1mm}
\begin{fmfgraph*}(11,15)
\fmfrightn{l}{2}
\fmfleft{r}
\fmf{scalar}{o,l1}
\fmf{scalar}{l2,o}
\fmf{boson}{o,r}
\fmfv{d.sh=di,d.f=full,d.si=2thick}{o}
\end{fmfgraph*}
}
\,\, +\,\,
\parbox{11mm}{
\setlength{\unitlength}{1mm}
\begin{fmfgraph*}(30,15)
\fmfrightn{l}{2}
\fmfleft{r}
\fmf{scalar}{ol,l1}
\fmf{scalar}{l2,ol}
\fmf{fermion,left=.5,tension=.5,label=$p$}{ol,or}
\fmf{heavy,left=.5,tension=.5,label=$\Lambda$}{or,ol}
\fmf{boson}{or,r}
\fmfv{d.sh=s,d.f=full,d.si=2thick}{ol}
\fmfv{d.sh=c,d.f=full,d.si=4thick}{or}
\end{fmfgraph*}
}\qquad\qquad\qquad .
\ee

$ $

\noindent
Here the small diamond symbolizes the bare coupling of  kaonic and
leptonic currents, the small box represents the matrix element of
the weak hadronic current between $\Lambda$ and proton-hole states
$
W_S^{\mu}= -\gamma^{\mu}\, (g^S_V+ g_A^S\, \gamma_5) $, the vector
coupling constant is $g_V^S=\sqrt{\frac32}$ and the axial coupling
constant is $g_A^S\approx 0.62\sqrt{\frac32}$. The $\Lb p$
interaction in the intermediate states given by Eq.~(\ref{tln}) is
absorbed into the dressed vertex of $KN\Lb$ interaction depicted by
the fat circle. The dashed lines symbolize an attached leptonic
weak current. From Eq.~(\ref{diagk}) we get
\be\label{G}
\sqrt{2}\, f_K\, \Gamma^{\mu}&=& \sqrt{2}\, f_K\, k^{\mu}
-\frac{g_A^S}{C_{KN\Lb}}\,  P_A^{\mu}\,,
\\ \nonumber
P_A^{\mu} &=&
i\,{C}_{KN\Lb}\, \tilde{C}_{KN\Lb}\, \int
\frac{d^4p}{(2\,\pi)^4}\, \Sp{\hk\,
\gamma_{5}\, \hat{G}_{\Lb}(p+k)\, \gamma^{\mu}\,\gamma_5\,
\hat{G}_N(p)}
\,.
\ee
Taking into account that $(P_A \cdot k)=\Pi^-_{\Lb}$, cf.
Eq.~(\ref{eq:loopint}), we obtain
\be\label{Gp}
\Gamma^{\mu}&=&
k^{\mu} \, \left(1- \frac{A}{\vk^2}\right) -\frac{\delta^{\mu 0}}
{\vk^2} \,  B \,,
\\ \nonumber\label{A} A(\om,\vk)&=& \,\Delta_S
\left[\om\, P^0_A(\om,\vk) -\Pi^-_{\Lb}(\om,\vk)\right]
\,,
\\ \nonumber
B(\om,\vk)&=& -\om\, A(\om,\vk)+\vk^2\,\Delta_S P^0_A(\om,\vk)
\,,\ee
where $\Delta_S=g_A^S/(\sqrt{2}\, C_{KN\Lb}\, f_K)$ is the
discrepancy of the Goldberger--Treiman relation, which is about
 67\%.

We will also consider  reactions with the leptonic current directly
attached to the nucleon--hyperon weak current. In this case, we
have to take into account the modification of the latter due to the
short-range correlation of nucleons and hyperons. We determine the
in-medium $\Lb$--$p$ weak current, $\tilde{W}^\mu_S$, by the
following diagram equation

$ $

\be\label{hypcurcor}
\tilde{W}^\mu_S=\parbox{20mm}{
\setlength{\unitlength}{1mm}
\begin{fmfgraph*}(20,15)
\fmfleftn{l}{2}
\fmfrightn{r}{2}
\fmf{heavy}{l2,o}
\fmf{fermion}{o,l1}
\fmf{scalar}{r1,o,r2}
\fmfv{d.sh=sq,d.f=full,d.si=4thick}{o}
\fmflabel{$\Lambda$}{l2}
\fmflabel{$p$}{l1}
\end{fmfgraph*}
}
=
\parbox{20mm}{
\setlength{\unitlength}{1mm}
\begin{fmfgraph*}(20,15)
\fmfleftn{l}{2}
\fmfrightn{r}{2}
\fmf{heavy}{l2,o}
\fmf{fermion}{o,l1}
\fmf{scalar}{r1,o,r2}
\fmfv{d.sh=sq,d.f=full,d.si=2thick}{o}
\fmflabel{$\Lambda$}{l2}
\fmflabel{$p$}{l1}
\end{fmfgraph*}
}
+
\parbox{20mm}{
\setlength{\unitlength}{1mm}
\begin{fmfgraph*}(40,15)
\fmfleftn{l}{2}
\fmfrightn{r}{2}
\fmf{heavy}{l2,ol}
\fmf{fermion}{ol,l1}
\fmf{scalar}{r1,or,r2}
\fmfv{d.sh=sq,d.f=shaded,d.si=4thick}{ol}
\fmfv{d.sh=sq,d.f=full,d.si=4thick}{or}
\fmf{fermion,left=.5,tension=.5,label=$p$}{or,ol}
\fmf{heavy,left=.5,tension=.5,label=$\Lambda$}{ol,or}
\fmflabel{$\Lambda$}{l2}
\fmflabel{$p$}{l1}
\end{fmfgraph*}
}\qquad\qquad\qquad .
\ee

$ $

\noindent
This current $\tilde{W}^\mu_S$ is irreducible with respect to
one-kaon exchange. The solution of Eq.~(\ref{hypcurcor}) with the
interaction (\ref{tln}) reads $
\tilde{W}_S^\mu=-\gamma^{\mu}\, \bigl[\gamma_\Lb(f_\Lb)\,g^S_V +
\gamma_\Lb(f'_\Lb)  g_A^S\, \gamma_5 \bigr] \,,
$
where the function $\gamma_\Lb$ is given by Eq.~(\ref{gamma}).

Provided with both weak interactions (\ref{diagk}) and
(\ref{hypcurcor}), and  with the kaon propagator in medium, we are
able to consider the neutrino induced reactions with associated
$S=-1$ strangeness production.

\section{Anti-neutrino scattering with strangeness production.}
The negative strangeness in a nucleus can be produced by neutrino
via the following reactions: (i) the neutrino decay
$\bar\nu_l\rightarrow  l^+ + K^-$ and (ii) the $\Lb$ production on
a nucleon $\bar{\nu}_l+p\rightarrow l^+ + \Lb$.  Other processes
with more particles in the initial and final states give smaller
contribution since their phase space volume is suppressed. As in
vacuum one may depict the processes (i)--(ii) by the following
diagrams

$ $

$$
\mbox{(i)}\qquad
\parbox{18mm}{
\setlength{\unitlength}{1mm}
\begin{fmfgraph*}(18,10)
\fmfleft{n}
\fmfright{k,l}
\fmf{scalar}{l,o,n}
\fmf{boson,width=1.5thick}{o,k}
\fmfforce{(0w,0.0h)}{n}
\fmfforce{(1.0w,0.0h)}{k}
\fmfforce{(1.0w,1.0h)}{l}
\fmfforce{(0.5w,0.0h)}{o}
\fmfv{d.sh=di,d.fi=full,d.si=4thick}{o}
\fmflabel{$\bar\nu_l$}{n}
\fmflabel{$l^+$}{l}
\fmflabel{$K^-$}{k}
\end{fmfgraph*}
}
%%%%%%%%%%%%%%%%%%
$$

$ $

$$
 \mbox{(ii$a$)}\qquad
%%%%%%%%%%%%%%%%%%
\parbox{30mm}{
\setlength{\unitlength}{1mm}
\begin{fmfgraph*}(30,10)
\fmfleftn{l}{2}
\fmfrightn{r}{2}
\fmf{fermion}{l1,o}
\fmf{heavy}{o,r1}
\fmf{scalar}{r2,o,l2}
\fmfforce{(0.0w,0.0h)}{l1}
\fmfforce{(0.0w,1.0h)}{l2}
\fmfforce{(0.5w,0.0h)}{o}
\fmfforce{(1.0w,0.0h)}{r1}
\fmfforce{(1.0w,1.0h)}{r2}
\fmfv{d.sh=sq,d.fi=full,d.si=4thick}{o}
\fmflabel{$p$}{l1}
\fmflabel{$\bar\nu_l$}{l2}
\fmflabel{$\Lb$}{r1}
\fmflabel{$l^+$}{r2}
\end{fmfgraph*}
}
\qquad +\quad \mbox{(ii$b$)}\qquad
\parbox{30mm}{
\setlength{\unitlength}{1mm}
\begin{fmfgraph*}(30,10)
\fmfleftn{l}{2}
\fmfrightn{r}{2}
\fmf{fermion}{l1,ok}
\fmf{heavy}{ok,r1}
\fmf{scalar}{r2,o,l2}
\fmf{boson,width=1.5thick}{ok,o}
\fmfforce{(0.0w,0.0h)}{l1}
\fmfforce{(0.0w,1.0h)}{l2}
\fmfforce{(0.5w,0.0h)}{ok}
\fmfforce{(0.5w,1.0h)}{o}
\fmfforce{(1.0w,0.0h)}{r1}
\fmfforce{(1.0w,1.0h)}{r2}
\fmfv{d.sh=di,d.fi=full,d.si=4thick}{o}
\fmfv{d.sh=c,d.fi=full,d.si=4thick}{ok}
\fmflabel{$p$}{l1}
\fmflabel{$\bar\nu_l$}{l2}
\fmflabel{$\Lb$}{r1}
\fmflabel{$l^+$}{r2}
\end{fmfgraph*}
}
%%%%%%%%%%%%%%%%%%%%
\qquad .
$$
%%%%%%%%%%%%%%%%%%%%%%%%%%%%%%%%%%%%%%%%%%%

$ $

\noindent
Taking into account in-medium effects, however, one uses the thick
wavy line for the in-medium $K^-$ meson and the fat vertices to
indicate in-medium renormalization given by Eqs~(\ref{correl}),
(\ref{diagk}) and (\ref{hypcurcor}).

Proceeding naively, one would calculate the amplitude of
strangeness production as the sum of the squared matrix elements of
reaction (i) and (ii), summing two contributions in (ii)
coherently. We show that this approach will immediately lead to
double counting.

Let us consider, first, the reaction (i), proposed by R.F. Sawyer
to test the kaon spectrum in medium. Utilizing the modified kaon
weak current we present the matrix element of the process
$\bar\nu_l\rightarrow K^- + l^+$ as
$${\cal M}_{K^-}=i\, G\, f_K \,\sin\theta_C\, (\Gamma(k)\cdot l)\,.$$
Here, $G$ stands for the Fermi constant of the weak interaction,
$\theta_C$ denotes the Cabbibo angle, and
$l_\mu=\bar{u}_{{\nu}}\,\gamma_\mu \, (1-\gamma_5)\,
u_{{l}}$ is the leptonic current. Carrying out summation of the
squared matrix element over the lepton spin and averaging over the
neutrino spin we obtain
\be \nonumber
V_{K^-}(E_\nu,\om,\vec{k\,})=\frac12\sum_{\rm spin} |{\cal
M}_{K^-}|^2= \frac12\, G^2\,f_K^2\,
\sin^2\theta_C\, \Gamma^{\alpha}(k)\, \Gamma^{\dagger\, \beta}(k)\,
\sum_{\rm spin} l_\alpha\, l_\beta^\dagger
\,.\ee
The summation over spins gives
\be\label{lepttens}
L^{\alpha\beta}= \sum_{\rm spin} l^\alpha\, l^{\dagger\,\beta}= 8\,
\biggl[p_l^\alpha\, p_\nu^\beta+ p_l^\beta\,
p_\nu^\alpha-g^{\alpha\beta}\,  (p_l\cdot p_\nu) -  i\,
\varepsilon^{\alpha\beta\gamma\delta}p_{l\, \gamma}\, p_{\nu\,
\delta}\biggr]\,,
\ee
and we find
\be\label{probampl}
V_{K^-}(E_\nu,\om,\vec{k\,})=
-4\,G^2\,f_K^2\,
\sin^2\theta_C\, \left[ m_l^2\, F_1(\om,\vec{k\,})-
2 \, F_2(\om,\vec{k\,},E_\nu) \right] \,,
\ee
where
$$
F_1(\om,\vk)= \left\{  \left|\om-\frac{\om\,A+B}{\vk^2}\right|^2
-\vk^2 \, \left| 1-\frac{A}{\vk^2}\right|^2 \right\}\,,
$$
$$
F_2(\om,\vk,E)= \frac{|B|^2}{\vk^4}\, \left\{ \left( E-\frac12 \,
\om \right)^2 - \frac14 \, \vk^2 \right\}\,.$$

\begin{figure}[h]
\begin{minipage}[b]{0.45\textwidth}
\epsfxsize=7cm
\bc\mbox{\epsfbox{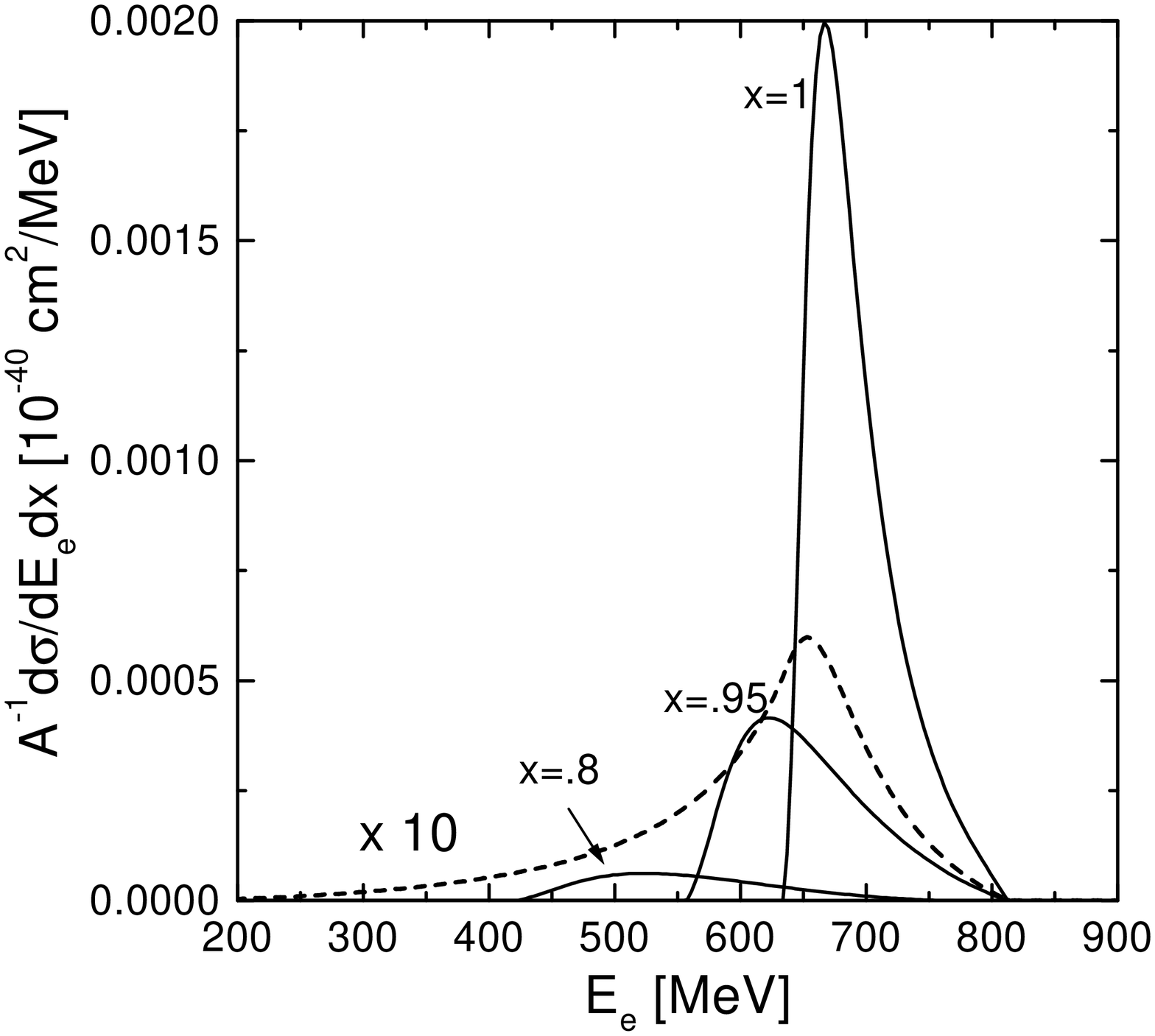}}\ec
  \end{minipage}
  \begin{minipage}[b]{0.45\textwidth}
\epsfxsize=7.2cm
\bc\mbox{\epsfbox{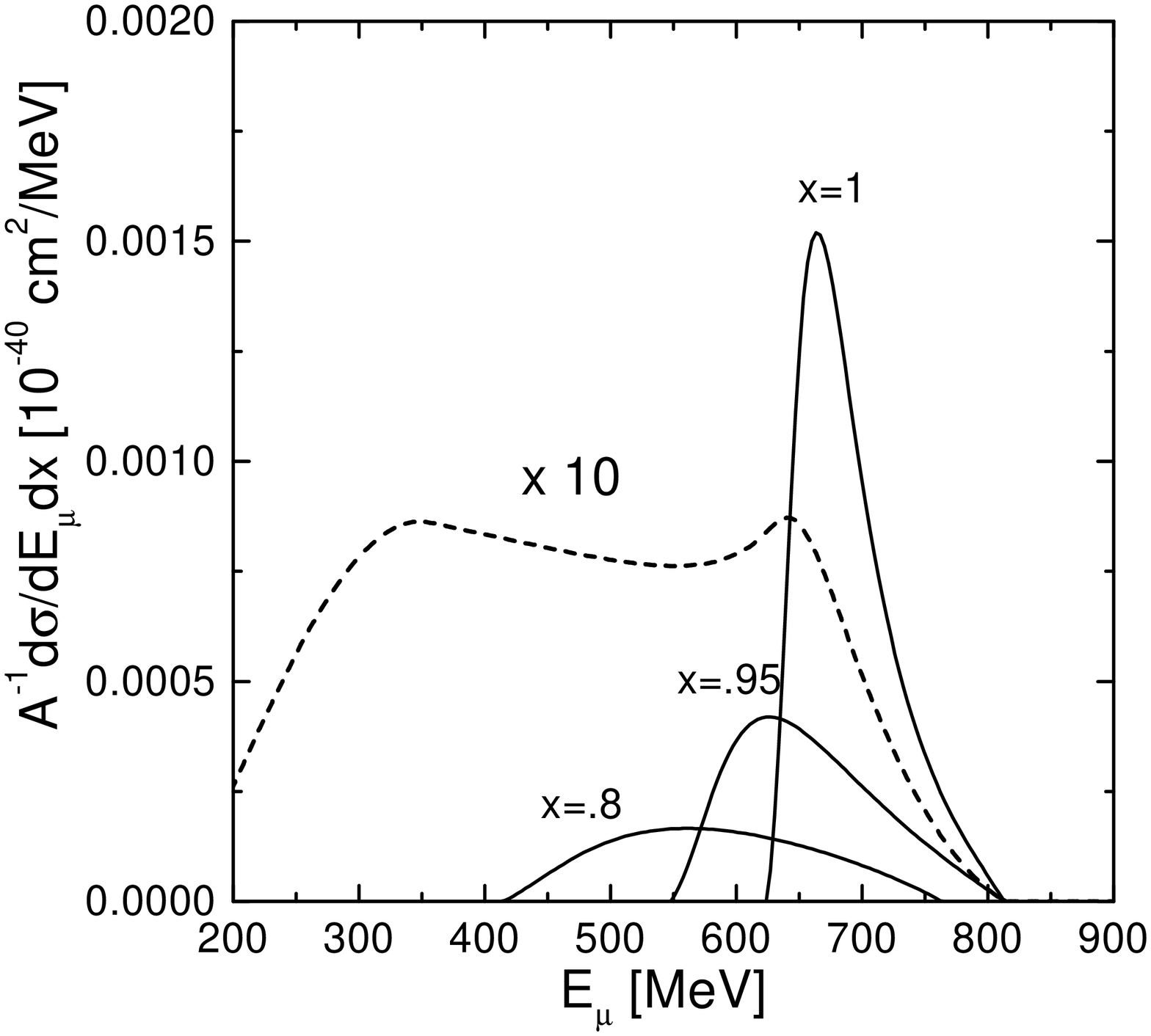}}\ec
\end{minipage}
\caption{Differential cross section (per particle) as the function
of the lepton energy for positron (left panel) and positive muon
(right panel) production in reaction $\bar\nu_l\rightarrow l^+
+K^-$ by anti-neutrinos scattering on a nucleus with beam energy
1~GeV. Solid lines are calculated for three values of the
scattering angle $\theta_l$ between an anti-neutrino and a lepton,
labeled by  the values of $x=\cos\theta_l$. Dashed lines  show the
cross section integrated over the lepton angle $\theta_l$.}
  \label{fig:emuspec}
\end{figure}

The differential production rate renders, then,
\be\label{diffwidth}
\frac{d{\cal W}_l^{\mbox{(i)}}}{dE_l\,dx\,dt}=
\frac{\sqrt{E_l^2-m_l^2}}{16\,\pi^2\,E_{\nu}} \,
\left[ -2\, \Im D_{K^-}^{R}(\bar{\om}_l,\bar{k}_l) \right] \, V_{K^-}
(E_{\nu},\bar{\om}_l,\bar{k}_l)\,.
\ee
Here $\bar{\om}_l=E_{\nu}-E_l$ is the kaon frequency for the
process with a given lepton in the final states $l=e^+,\mu^+$ ,
and $\bar{k}_l=
\sqrt{E_{\nu}^2+E_l^2-m_l^2-2\,x_l E_{\nu}\, \sqrt{E_l^2-m_l^2}}
$ is the corresponding kaon momentum, where $x_l=\cos\theta_l\,\,$
for $\theta_l$, being the angle between an incoming anti-neutrino
and an outgoing lepton. Considering the nucleus with a weight $A$
to be a uniform sphere of the radius $R=r_0\, A^{1/3}$ (where
$r_0\simeq 1.2$~fm), we present the differential cross section of the
positive lepton production as
\be\label{nucros}
\frac{d\sigma_l^{\mbox{(i)}}}{dE_{l}\,dx}=
2\,\pi\,r_0^3\, A \,
\frac{d{\cal W}_l^{\mbox{(i)}}}{dE_l\,dx\,dt}\,.
\ee
In Fig.~\ref{fig:emuspec} we show the differential cross sections
per $A$ of the $e^+$ (left panel) and $\mu^+$ (right panel)
production by a neutrino with energy $E_{\nu}=1$~GeV. We
observe that the cross sections are strongly peaked as the function
of lepton energy for small lepton scattering angles, $x > 0.95$. At
larger angles $0.8 < x < 0.95$, the cross sections decrease rapidly
by factor 5 for $e^+$ and 3 for $\mu^+$, and are shifted to smaller
lepton energies. At angles corresponding to $x < 0.8$, the positron
production cross section decreases further and becomes almost
negligible, whereas for positive muons the cross section decreases
moderately. Due to this, the angular integrated cross section of
muons is larger than that of positrons, especially at smaller
lepton energies. Having integrated over the lepton energy, we
obtain for the total cross sections of the $e^+(\mu^+)$ production
$$ A^{-1}\,
\sigma(\bar\nu_e\rightarrow K^-+e^+)\approx 1 \times 10^{-42}\, {\rm cm}^2
\, , \quad A^{-1}\,
\sigma(\bar\nu_\mu\rightarrow K^-+\mu^+)\approx 4 \times
10^{-42}\, {\rm cm}^2\,.
$$
We note that the $\mu^+$ production cross section is larger than
positron one by factor $\approx 4$, only. This is in contrast to
expectation based upon the vacuum branching rations of a kaon decay
$\Gamma(K^-\rightarrow e^-+\nu_e)/\Gamma(K^-\rightarrow
\mu^-+\nu_\mu)\approx 2.5\times 10^{-5}$. For the bare weak interaction
the squared matrix element of the reaction $\bar\nu_l\rightarrow
K^- +l^+$ would be $|{\cal M}|^2\propto m_l^2[m_l^2-(k\cdot k)]/2$,
that explains a strong enhancement of muon processes in vacuum
compared with the positron ones. In medium the weak kaon current is
dramatically modified due to the mixture of kaons with the
$\Lb$-particles--proton hole states carrying the same quantum
numbers, see the second diagram in Eq.~(\ref{diagk}). As the
result, the squared matrix element of the reaction (\ref{probampl})
does not possess  strong dependence on the lepton mass. Therefore,
in medium squared matrix elements for positrons and muons turn out
to be of the same order of magnitude.

Please notice that the process considered here, following the
arguments of R.F. Sawyer, probes the kaon spectral density in the
whole frequency--momentum region rather than nearby the kaon branch
as it was considered in Ref.~\cite{sawyerkaon}. Indeed, the $K^-$
spectral density in Eq.~(\ref{diffwidth}) can be split as following
\be\label{spd}
-2\Im D_{K^-}^R(\om,k)&=&2\,\Im\Pi_\Lb(\om,k)\, |D_{K^-}^R(\om,k)|^2
+2\,\delta\Im \Pi_{K^-}(\om,k) \, | D_{K^-}^R(\om,k)|^2 \,.
\ee
 The first term
in the decomposition (\ref{spd}) corresponds to the low-energy
kaonic states in the $\Lb$--proton-hole continuum, whereas the
second one is related to the contribution of other kaon dissipation
processes. In our approximation for the s-wave $KN$ interaction the
residual part of the spectral density is $\delta$-function-like
\be\label{spdres}
2\,\delta\Im \Pi_{K^-}(\om,k) \,  |D_{K^-}^R(\om,k)|^2 =2\, \pi\,
\delta\bigl(\om^2-k^2-m_K^2-\Re\Pi_{K^-}(\om,k)\bigl)\,
\theta(\om-\om^+_{p\Lb}(k))\,,
\ee
where $\theta(x)$ is the Heaviside's step function.
 In Ref.~\cite{sawyerkaon} only the second term in Eq.~(\ref{spd}) was
considered, in which the $\delta$ function was smoothed by a
constant width. In our case this term starts to contribute only
when the kaon momentum exceeds the critical value $k_c$, which is
the solution of the equation
$m_K^2+\Re\Pi_{K^-}(\om=k_c,|\vec{k}|=k_c)=0$. For our polarization
operator, constrained by the low-energy theorems, the value
$k_c=2570$~MeV is substantially larger than that ($\sim 1600$~MeV)
obtained in Ref.~\cite{sawyerkaon}. (We recover the latter value
putting $\lb=0$.) Therefore, it seems that the region of the kaon
spectral density considered in Ref.~\cite{sawyerkaon} is unlikely
to be probed by anti-neutrinos with energies less than the
threshold value $E_\nu^{\rm thr}\simeq 2570$~MeV.

On the other hand, we have seen that the first term in the kaon
spectral density  contributes at much smaller neutrino energies.
However, in this case we deal with the kaon which, being produced,
decays into the $\Lb$ particle and the proton-hole, i.e., this
process occurs exactly in the same neutrino-energy region, where
the process $\bar{\nu}_{l}+p\rightarrow \Lb + l^+$ does. This
invites us to investigate the probability of the latter in more
detail.

According to the diagrams (ii) above, the matrix element of the
reaction $\bar{\nu}_l+p\rightarrow
\Lb + l^+$ can be written as
$${\cal
M}_{\Lb}^{\mbox{(ii)}}= {\cal M}_\Lb^{\mbox{(ii$a$)}}+{\cal
M}_\Lb^{\mbox{(ii$b$)}}
=
\frac{1}{\sqrt{2}}\,G\, \sin\theta_C \,l_\mu\,\bar{u}_\Lb (
\tilde{W}_S^\mu+\sqrt{2}\, f_K\, \tilde{C}_{KN\Lb}\,\Gamma^\mu\,\hat{k}\,
\gamma_5\,D_{K^-}\, )\, u_p\,.$$
For the squared, spin averaged matrix element we obtain
\be\label{mlamb}
 \frac12\sum_{\rm spin}
|{\cal M}_\Lb|^2=\frac12\sum_{\rm spin} |{\cal M}_\Lb^{\mbox{(ii$a$)}}|^2
+\frac12\sum_{\rm spin} 2\, \Re\left\{ {\cal
  M}_\Lb^{\mbox{(ii$b$)}}{\cal M}_\Lb^{\dagger\,\mbox{(ii$a$)}}\right\} +
\frac12\sum_{\rm spin}  |{\cal M}_\Lb^{\mbox{(ii$b$)}}|^2
\ee
with
\be \label{miia}
\frac12\sum_{\rm spin}
|{\cal M}_\Lb^{\mbox{(ii$a$)}}|^2= \frac12 \, G^2\, \sin^2\theta_C\,
\Sp{ \tilde{W}^\mu_S\, (\hat{p}_\Lb+m^*_\Lb)\, \tilde{W}^{\dagger\,\nu}_S
(\hat{p}_p+\mn^*)}\, L_{\mu\nu} \,,
\ee\be \nonumber
\frac12\sum_{\rm spin} 2\, \Re\left\{ {\cal
  M}_\Lb^{\mbox{(ii$b$)}}{\cal M}_\Lb^{\dagger\,(iia)}\right\}
=G^2\,\sin^2\theta_C\,f_K
\ee\be \label{miiab}
\times
2\, \Re \left[\tilde{C}_{KN\Lb}^\dagger\,
  D_{K^-}^{R\,\dagger}\,\Gamma^{\dagger\, \mu}\,
 \Sp{\hk\, \gamma_{5}\, (\hat{p}_\Lb+m^*_\Lb)\, \tilde{W}^\nu_S
(\hat{p}_p+\mn^*)}\right]\, L_{\mu\nu}\,,
\ee\be \label{miib}
\frac12\sum_{\rm spin}
|{\cal M}_\Lb^{\mbox{(ii$b$)}}|^2
=V_{K^-}(E_\nu,\om,k)\,|\tilde{C}_{KN\Lb}|^2\,|D_{K^-}^R|^2
\Sp{\hk\, \gamma_{5}\,
  (\hat{p}_\Lb+m^*_\Lb)\,\hk\, \gamma_{5}\,  (\hat{p}_p+\mn^*)}
\,.\ee
Here $L_{\mu\nu}$ is the leptonic tensor (\ref{lepttens}), and
$p_\Lb$  and $p_p$ are momenta of the $\Lb$ particle and the
proton, respectively. Utilizing kinematics of the reaction, the
first term in Eq.~(\ref{mlamb}) renders
\be\nonumber
\frac12\sum_{\rm spin}
|{\cal M}_\Lb^{\mbox{(ii$a$)}}|^2&=&
V_{\Lb}((p_p\cdot p_\nu),\om,\vec k)\\ \nonumber
&=& G^2\,\sin^2\theta_C\,
\Biggl\{ 4\,[(k\cdot k)-\Delta]\,[(k\cdot k) -m_l^2]
\\ \nonumber
&\times&  \biggl[ (g^{S}_V)^2 |\gamma_\Lb (f_\Lb)|^2
-g^{S}_A\,g^{S}_V\,\Re[\gamma_\Lb(f_\Lb)\gamma_\Lb^\dagger(f'_\Lb)]
+(g^{S}_A)^2|\gamma_\Lb(f'_\Lb)|^2\biggr]
\\ \nonumber
&+& 16\,\biggl[(g^{S}_V)^2 |\gamma_\Lb (f_\Lb)|^2 + (g^{S}_A)^2
|\gamma_\Lb (f'_\Lb)|^2 \biggr]
\\ \nonumber
&\times&\biggl(2\,(p_p\cdot p_\nu)^2+(p_p\cdot p_\nu)\,[(k\cdot
k)-\Delta-m_l^2]\biggr)
\\ \nonumber
&-&16\,g^{S}_A\,g^{S}_V\,\Re[\gamma_\Lb(f_\Lb)\gamma_\Lb^\dagger(f'_\Lb)]
\,(p_p\cdot p_\nu)\,(k\cdot k)
\\  \label{probamplLb}
&+&8\,\biggr[(g^{S}_V)^2 |\gamma_\Lb (f_\Lb)|^2 - (g^{S}_A)^2
|\gamma_\Lb (f'_\Lb)|^2 \biggl] \mn^*\,m_\Lb^*\,[m_l^2-(k\cdot k)]
\Biggr\} \,,\ee
where  $(k\cdot k)=\om^2-\vec{k\,}^2$,
$\Delta=m_\Lb^{*\,2}-\mn^{*\,2}$, and $p_\nu$ and $p_p$ are momenta
of an anti-neutrino and a proton, respectively. In
Eqs~(\ref{miiab}) and (\ref{miib}) we recognize the traces
appearing in Eqs.~(\ref{G}) and (\ref{eq:loopint}), that allow us
to calculate them with ease.

After integrating over the phase-space volume we obtain the
differential rate of the reaction $\bar{\nu}_{l}+p\rightarrow
\Lb + l^+$ as
\be\label{diffwidthLb}
\frac{d{\cal W}_l^{\mbox{(ii)}}}{dE_l\,dx\,dt}&=&
\frac{d{\cal W}_l^{\mbox{(ii$a$)}}}{dE_l\,dx\,dt} +
\frac{d{\cal W}_l^{\mbox{(ii$ab$)}}}{dE_l\,dx\,dt} +
\frac{d{\cal W}_l^{\mbox{(ii$b$)}}}{dE_l\,dx\,dt}
\,.\ee
Three terms here are the contributions from diagram (ii$a$),
interference term between diagrams (ii$a$) and (ii$b$), and diagram
(ii$b$), respectively. The first term yields
\be \label{iia}
\frac{d{\cal W}_l^{\mbox{(ii$a$)}}}{dE_l\,dx\,dt}=
\frac{\sqrt{E_l^2-m_l^2}}{16\,\pi^2\,E_{\nu}} \,2\,\Im \left[(-i)
\int \frac{d^4 p}{(2\pi)^4}G_N(p)\, G_\Lb(p+\bar{k}_l)\,
V_\Lb((p_\nu\cdot p),\bar{\om}_l,\bar{k}_l) \right]
\,,\ee
where the frequency $\bar\om_l$ and the momentum $\bar k_l$ are
defined as in Eq.~(\ref{diffwidth}). In the second term we use
$\Delta_S \,\Im P_A^\mu=-\Im\Gamma^\mu$ and separate explicitly the
real and imaginary parts of the kaon propagator. Then, the second
term reads
\be \nonumber
\frac{d{\cal W}_l^{\mbox{(ii$ab$)}}}{dE_l\,dx\,dt} =
\frac{\sqrt{E_l^2-m_l^2}}{16\,\pi^2\,E_{\nu}}
\Bigl\{2\,\Re D_{K^-}^R(\bar{\om}_l,\bar k_l)\,
V_K^{(1)}(E_\nu,\bar\om_l,\bar k_l)
\\ \label{iiab} - 2\,\Im D_{K^-}^R
(\bar\om_l,\bar k_k)\, V_K^{(2)}(E_\nu,\bar\om_l,\bar k_l)
\Bigr\}
\,,\ee
where
\be\nonumber
V_K^{(1)}(E_\nu,\om, k) &=& 2\, G^2\, \sin^2\, \theta_C
\,f_K^2\, (\Re\Gamma^\mu\,\Im\Gamma^{\dagger \nu} \, L_{\mu\nu})\,,
\\ \nonumber
V_K^{(2)}(E_\nu,\om, k) &=& 2\, G^2\, \sin^2\, \theta_C
\,f_K^2\, (\Im\Gamma^\mu\,\Im\Gamma^{\dagger \nu} \, L_{\mu\nu})
\,.\ee
The last term in Eq.~(\ref{diffwidthLb}) renders
\be \label{iib}
\frac{d{\cal W}_l^{\mbox{(ii$b$)}}}{dE_l\,dx\,dt}
=\frac{\sqrt{E_l^2-m_l^2}}{16\,\pi^2\,E_{\nu}}\,
\left[-2\, \Im\Pi_{\Lb}(\bar{\om}_l,\bar{k}_l)\,
|D_{K^-}^R(\bar{\om}_l,\bar{k}_l)|^2\right]  V_{K^-}(E_\nu,\om,k)
\,.\ee
Comparing this expression with Eqs~(\ref{diffwidth}) and
(\ref{spd}) we observe that exactly the same term has been already
taken into account in Eq.~(\ref{diffwidth}) with  the first term of
the kaon spectral density (\ref{spd}). This demonstrates the
mentioned problem of  double counting, which appears if one blindly
includes medium effects in Feynman diagrams.

The differential cross section of the $l^+$ production in the
reaction $\bar\nu_l + p\to \Lb +l^+$ can be calculated using
Eq.~(\ref{nucros})
\be\label{nucroslb}
\frac{d\sigma_l^{\mbox{(ii)}}}{dE_{l}\,dx}=
2\,\pi\,r_0^3\, A \, \left\{
\frac{d{\cal W}_l^{\mbox{(ii$a$)}}}{dE_l\,dx\,dt} +
\frac{d{\cal W}_l^{\mbox{(ii$ab$)}}}{dE_l\,dx\,dt}\right\}\,.
\ee
Here we droped the term (\ref{iib}) included in our analysis of
process (i).

\begin{figure}[h]
\epsfxsize=7cm
\bc\mbox{\epsfbox{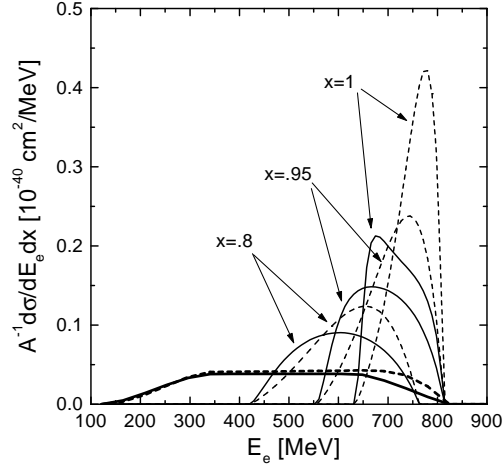}}\ec
\caption{Differential cross section (per particle) of positrons
produced in reaction $\bar \nu_l+p\rightarrow l^+ +\Lb$ by
anti-neutrino of beam energy 1~GeV. Thin solid lines correspond to
calculations with in-medium vertex renormalization. Thin dashed
lines show the result without inclusion of  short-range $\Lb N$
correlations. Thick solid and dashed lines depict the cross
sections integrated over the lepton angle $\theta_l$ with and
without account for vertex renormalization.}
\label{fig:lbspec}
\end{figure}

Fig.~\ref{fig:lbspec} shows the results for the anti-neutrino
energy of 1~GeV. We find that the difference between the positron
and muon reactions is very small, therefore in
Fig.~\ref{fig:lbspec} we show the result for positron production
only. The solid and dashed lines are related to calculations
without and with account for the in-medium renormalization of the
weak interaction in process (ii$a$), i.e., in the first term in
Eq.~(\ref{nucroslb}). To be specific, in our calculation we put
$f_\Lb=f'_\Lb$. In Fig.~\ref{fig:lbspec} we see that the
short-range $\Lb N$ correlations (factors $\gamma_\Lb$ in
Eq.~(\ref{probamplLb})) suppress the cross section of the reaction
(ii$a$) and change the shape of the lepton spectrum.

The contribution from the interference terms between processes
(ii$a$) and (ii$b$) is found to be very small and is not
distinguishable on the scale of Fig.~\ref{fig:lbspec}.

In Fig.~\ref{fig:lbspec} we observe that the positron production
cross section decreases monotonically with the increasing positron
scattering angle. The angular integrated cross section remains
almost constant in a wide interval of the positron energy. The
total cross section of the $\Lb$ production on a nucleus is
$$A^{-1}\,
\sigma(\bar\nu_e+p\to e^++\Lb)\approx
A^{-1}\,\sigma (\bar\nu_\mu+p\to \mu^+ + \Lb)
\approx 2\times 10^{-39} \,{\rm cm}^2\,.
$$
In the total cross section the correlations result in a decrease
of $\sim$10\%.

As we can see,  reaction $\bar{\nu}_{l} +p
\rightarrow l^+ +\Lb$ gives the main contribution to the
strangeness production by anti-neutrino on a nucleus\footnote{The
cross sections in Fig.~\ref{fig:emuspec} and those of
Ref.~\cite{sawyerkaon} are of the same order of magnitude despite
the different energies used.}. Besides, this process occurs in the same
kinematic region as the reaction $\bar{\nu}_{l}
\rightarrow l^+ +K^-$. Therefore, one needs more peculiar analysis,
 in order to separate the contribution of the $K^-$ channel.

In principle, one can suggest to observe directly $K^-$ mesons
produced by an anti-neutrino on a nucleus. However, the kaons
produced in reaction (i) are too far off mass-shell to lap from an
in-medium state to a vacuum one. They have to gather  the energy in
the sequence of proceeding rescattering processes. Additionally
free kaons can be produced in the two-step processes with a $\Lb$
decay, e.g., $\bar\nu_l+p\to l^+ +\Lb \to l^+ +p+ K^-$. This is a
surface reaction, since the $K^-$ has rather short mean free path
in nuclear matter. Thereby, the yield of this reaction is
suppressed. The similar process with a direct pion production has
been considered in Ref.~\cite{oset97}.

\section{optical-theorem formalism for neutrino scattering}
The example, considered above, demonstrates clearly that naive
account of in-medium effects could lead to  double counting. In the
particular simplified case it was rather easy to resolve the
problem. In general case, with account for more in-medium degrees
of freedom coupled to the neutrino--lepton weak current, the double
counting problem becomes very serious. Therefore, one needs an
approach, within which such a problem does not exists or can be
easily resolved. In Refs.~\cite{vossen,kvaop} it was shown that the
formalism of the optical theorem formulated in terms of
non-equilibrium Green's functions allows to avoid the double
counting problem.

Applying this approach to the anti-neutrino--nucleus scattering we
 can express the transition probability between the initial state with an
anti-neutrino $\bar\nu_l$ and the final state with a positive
lepton $l^+$ in terms of an evolution operator $S$ as follows
\be\label{optfirst}
\frac{d{\cal W}^{\rm  tot}_{\bar\nu\to l^+}}{d t}=
\frac{d p_l^3}{(2\pi)^3\,4\,E_{{\nu}}\,E_l}
\,\sum_{\{X\}}\overline{<\bar\nu_l|\, S^\dagger\, |l^+ + X>\,<l^++X|\,
S\, |\bar\nu_l>}\,,
\ee
where we write explicitly the phase-space volume of initial
($\bar\nu_l$) and final ($l^+$) states. The bar denotes statistical averaging.
The summation goes over complete set of all possible intermediate
states $\{X\}$ constrained by the energy--momentum conservation
law.

Making use of the smallness of the Fermi weak-interaction constant
$G$, we can take into account processes in first order in $G$.
Then, we expand the evolution operator $S$ as
\be\label{Smatrix}
S\approx 1- i\, \intop_{-\infty}^{+\infty} T\,\bigl\{V_W(x)\,
S_{\rm nucl} (x)\bigr\} d x_0
\,,\ee
where $V_W$ is the Hamiltonian of the weak interaction,
$V_W=\frac{G}{\sqrt{2}}\,l_\mu\,(W_S^\mu+J^\mu_K)$, taken in
Eq.~(\ref{Smatrix}) in the interaction representation, and $S_{\rm
nucl}$ is the part of the $S$ matrix corresponding to the nuclear
interaction, and $T\{...\}$ stands for the chronological  ordering
operator. After substitution of the $S$ matrix (\ref{Smatrix}) into
Eq.~(\ref{optfirst}) and averaging over the arbitrary
non-equilibrium state of a nuclear system, there appear
chronologically, anti-chronologically ordered exact Green's
functions, denoted as $G^{--}$ and $G^{++}$ respectively, and
disordered Green's functions $G^{+-}$ and $G^{-+}$. The latter ones
are related to Wigner's densities.

In  graphical form the general expression for the probability of
positive lepton production by an anti-neutrino is determined by the
diagram

$ $

$$ \parbox{40mm}{\setlength{\unitlength}{1mm}
\begin{fmfgraph*}(40,15)
\fmfleftn{l}{2}
\fmfrightn{r}{2}
\fmfpoly{hatched,pull=1.4,smooth}{ord,oru,olu,old}
\fmfforce{(0.3w,.8h)}{olu}
\fmfforce{(0.7w,.8h)}{oru}
\fmfforce{(0.3w,0.2h)}{old}
\fmfforce{(0.7w,0.2h)}{ord}
\fmfforce{(0.25w,0.5h)}{ol}
\fmfforce{(0.75w,0.5h)}{or}
\fmf{scalar}{l1,ol}
\fmf{scalar}{ol,l2}
\fmf{scalar}{r2,or}
\fmf{scalar}{or,r1}
\fmflabel{$\bar\nu_l$}{l1}
\fmflabel{$l^+$}{l2}
\fmflabel{$\bar\nu_l$}{r2}
\fmflabel{$l^+$}{r1}
\fmfv{l=$+$,l.a=180,l.d=3thick}{ol}
\fmfv{l=$-$,l.a=0,l.d=3thick}{or}
\end{fmfgraph*}}
\,\,\, ,$$

$ $

\noindent
which  represents the sum of all closed diagrams containing at
least one ($+-$) line.  The contributions of specific processes
contained in a closed diagram  can be made visible by cutting the
diagram over the ($+-$), ($-+$) lines, corresponding to exact
$G^{+-}$ and $G^{-+}$ Green's functions.

The various contributions from $\{X\}$ can be classified according
to global characteristics, such as strangeness, parity etc. Then,
we can write, e.g.,
\be\nonumber
\frac{d{\cal W}^{\rm  tot}_{\bar\nu\to l^+}}{d t}&=&
\frac{d{\cal W}^{\Delta S=0}_{\bar\nu\to {l}^+}}{d t} +\frac{d{\cal
W}^{\Delta S=-1}_{\bar\nu\to {l}^+}}{d t} +\dots
\\ \nonumber \\ \nonumber\\
\label{opticgen}
&=&\frac{d^3 p_l}{(2\pi)^3\,4\,E_{{\nu}}\,E_l}
\left(\,\,\,
\parbox{40mm}{
\setlength{\unitlength}{1mm}
\begin{fmfgraph*}(40,15)
\fmfleftn{l}{2}
\fmfrightn{r}{2}
\fmfpoly{empty,pull=1.4,smooth,label=
$\Delta S=0$ }{ord,oru,olu,old}
\fmfforce{(0.3w,.8h)}{olu}
\fmfforce{(0.7w,.8h)}{oru}
\fmfforce{(0.3w,0.2h)}{old}
\fmfforce{(0.7w,0.2h)}{ord}
\fmfforce{(0.25w,0.5h)}{ol}
\fmfforce{(0.75w,0.5h)}{or}
\fmf{scalar}{l1,ol}
\fmf{scalar}{ol,l2}
\fmf{scalar}{r2,or}
\fmf{scalar}{or,r1}
\fmflabel{$\bar\nu_l$}{l1}
\fmflabel{$l^+$}{l2}
\fmflabel{$\bar\nu_l$}{r2}
\fmflabel{$l^+$}{r1}
\fmfv{l=$+$,l.a=180,l.d=3thick}{ol}
\fmfv{l=$-$,l.a=0,l.d=3thick}{or}
\end{fmfgraph*}
}
\,\,+\,\,
\parbox{40mm}{
\setlength{\unitlength}{1mm}
\begin{fmfgraph*}(40,15)
\fmfleftn{l}{2}
\fmfrightn{r}{2}
\fmfpoly{empty,pull=1.4,smooth,label=
$\Delta S= -1$ }{ord,oru,olu,old}
\fmfforce{(0.3w,.8h)}{olu}
\fmfforce{(0.7w,.8h)}{oru}
\fmfforce{(0.3w,0.2h)}{old}
\fmfforce{(0.7w,0.2h)}{ord}
\fmfforce{(0.25w,0.5h)}{ol}
\fmfforce{(0.75w,0.5h)}{or}
\fmf{scalar}{l1,ol}
\fmf{scalar}{ol,l2}
\fmf{scalar}{r2,or}
\fmf{scalar}{or,r1}
\fmflabel{$\bar\nu_l$}{l1}
\fmflabel{$l^+$}{l2}
\fmflabel{$\bar\nu_l$}{r2}
\fmflabel{$l^+$}{r1}
\fmfv{l=$+$,l.a=180,l.d=3thick}{ol}
\fmfv{l=$-$,l.a=0,l.d=3thick}{or}
\end{fmfgraph*}
}
\,\,\,\dots\right)\,\, .
\ee

$ $

\noindent
The first term represents all processes with the total strangeness
$0$ in the intermediate states. The second term contains the
processes with the total strangeness $-1$. Ellipses symbolize all
other processes. In Ref.~\cite{vossen} it was shown that each blob
in Eq.~(\ref{opticgen}) can be considered as a propagation of some
quanta of the in-medium interaction with certain quantum numbers.
We illustrate it with the example of strangeness production,
considered explicitly in the previous section.

\subsection{Strange channel}
Restricting our consideration to processes (ii$a$) and (ii$b$), we
decompose the second blob in Eq.~(\ref{opticgen}) as

$ $

\be\label{opts}
\parbox{40mm}{
\setlength{\unitlength}{1mm}
\begin{fmfgraph*}(40,15)
\fmfleftn{l}{2}
\fmfrightn{r}{2}
\fmfpoly{empty,pull=1.4,smooth,label=
$\Delta S= -1$ }{ord,oru,olu,old}
\fmfforce{(0.3w,.8h)}{olu}
\fmfforce{(0.7w,.8h)}{oru}
\fmfforce{(0.3w,0.2h)}{old}
\fmfforce{(0.7w,0.2h)}{ord}
\fmfforce{(0.25w,0.5h)}{ol}
\fmfforce{(0.75w,0.5h)}{or}
\fmf{scalar}{l1,ol}
\fmf{scalar}{ol,l2}
\fmf{scalar}{r2,or}
\fmf{scalar}{or,r1}
\fmflabel{$\bar\nu_l$}{l1}
\fmflabel{$l^+$}{l2}
\fmflabel{$\bar\nu_l$}{r2}
\fmflabel{$l^+$}{r1}
\fmfv{l=$+$,l.a=180,l.d=3thick}{ol}
\fmfv{l=$-$,l.a=0,l.d=3thick}{or}
\end{fmfgraph*}
}
\,\,\,=\,\,\,
\setlength{\unitlength}{1mm}
\parbox{40mm}{\begin{fmfgraph*}(40,15)
%\fmfpen{thick}
\fmfleftn{l}{2}
\fmfrightn{r}{2}
\fmf{scalar}{l1,ol,l2}
\fmf{scalar}{r2,or,r1}
\fmfforce{(0.28w,0.5h)}{ol}
\fmfforce{(0.72w,0.5h)}{or}
\fmf{dbl_dots,label=$K^-$}{ol,or}
\fmflabel{$\bar\nu_l$}{l1}
\fmflabel{$l^+$}{l2}
\fmflabel{$\bar\nu_l$}{r2}
\fmflabel{$l^+$}{r1}
\fmfv{l=$+$,l.a=180,l.d=3thick}{ol}
\fmfv{l=$-$,l.a=0,l.d=3thick}{or}
\fmfv{d.sh=di,d.fi=full,d.si=2thick}{ol,or}
\end{fmfgraph*}}
\,\,\, + \,\,\,
\setlength{\unitlength}{1mm}
\parbox{40mm}{\begin{fmfgraph*}(40,15)
\fmfleftn{l}{2}
\fmfrightn{r}{2}
\fmf{scalar}{l1,ol,l2}
\fmf{scalar}{r2,or,r1}
\fmfforce{(0.2w,0.5h)}{ol}
\fmfforce{(0.8w,0.5h)}{or}
\fmfpoly{shade}{old,olu,ol}
\fmfpoly{shade}{or,oru,ord}
\fmfforce{(0.35w,0.8h)}{olu}
\fmfforce{(0.35w,0.2h)}{old}
\fmfforce{(0.65w,0.8h)}{oru}
\fmfforce{(0.65w,0.2h)}{ord}
\fmf{heavy,left=.4,tensio=.5}{olu,oru}
\fmf{fermion,left=.4,tension=.5}{ord,old}
\fmflabel{$\bar\nu_l$}{l1}
\fmflabel{$l^+$}{l2}
\fmflabel{$\bar\nu_l$}{r2}
\fmflabel{$l^+$}{r1}
\fmfv{l=$+$,l.a=180,l.d=3thick}{ol}
\fmfv{l=$-$,l.a=0,l.d=3thick}{or}
\end{fmfgraph*}} \,\,\, \dots
\,\,\, .\ee

$ $

\noindent
Ellipses symbolize other more complicated processes with more
($+-$), ($-+$) lines  in the intermediate states, which within the
Feynman diagram formalism would be depicted by the diagrams with
more particles in initial and final states. The contributions of
such processes are suppressed due to the smaller phase-space
volume. By this reason we drop them.

The decomposition (\ref{opts}) is done according to the following
principles: We separate two channels with strangeness exchange via
a kaon (the first diagram) and a $\Lb$--proton-hole (the second
diagram). The kaon exchange in the first diagram has to be
irreducible with respect to the $\Lb$--proton-hole. Therefore, the
dotted line symbolizes an in-medium kaon dressed by the s-wave and
residual parts of the kaon polarization operator only. In diagrams
it can be shown as follows
\be\label{ksdressed}
\setlength{\unitlength}{1mm}
\parbox{10mm}{
\begin{fmfgraph}(10,6)
\fmfforce{(0.0w,0.5h)}{l}
\fmfforce{(1.0w,0.5h)}{r}
\fmf{dbl_dots}{l,r}
\end{fmfgraph}}
\,\,\,=\,\,\,
\parbox{10mm}{
\begin{fmfgraph}(10,6)
\fmfforce{(0.0w,0.5h)}{l}
\fmfforce{(1.0w,0.5h)}{r}
\fmf{boson}{l,r}
\end{fmfgraph}}
\,\,\,+\,\,\,
\parbox{30mm}{
\begin{fmfgraph*}(30,10)
\fmfforce{(0.0w,0.5h)}{l}
\fmfforce{(1.0w,0.5h)}{r}
\fmfpoly{empty,label=$\Pi_S$}{rd,ru,lu,ld}
\fmfforce{(0.3w,0.0h)}{ld}
\fmfforce{(0.3w,1.0h)}{lu}
\fmfforce{(0.7w,0.0h)}{rd}
\fmfforce{(0.7w,1.0h)}{ru}
\fmfforce{(0.3w,0.5h)}{kl}
\fmfforce{(0.7w,0.5h)}{kr}
\fmf{boson}{l,kl}
\fmf{dbl_dots}{kr,r}
\end{fmfgraph*}}
\,\,\,+\,\,\,
\parbox{30mm}{
\begin{fmfgraph*}(30,10)
\fmfforce{(0.0w,0.5h)}{l}
\fmfforce{(1.0w,0.5h)}{r}
\fmfpoly{empty,label=$\Pi_{\rm res}$}{rd,ru,lu,ld}
\fmfforce{(0.3w,0.0h)}{ld}
\fmfforce{(0.3w,1.0h)}{lu}
\fmfforce{(0.7w,0.0h)}{rd}
\fmfforce{(0.7w,1.0h)}{ru}
\fmfforce{(0.3w,0.5h)}{kl}
\fmfforce{(0.7w,0.5h)}{kr}
\fmf{boson}{l,kl}
\fmf{dbl_dots}{kr,r}
\end{fmfgraph*}}
\,\,\, .\ee

\noindent
The shaded vertex in the second diagram in (\ref{opts}) is
irreducible with respect to the ($+-$) and ($-+$) kaon line and the
($+-$) and ($-+$) $\Lb$--proton-hole lines. This means it contains
only the lines of a given sign, all ($--$) or ($++$). Thereupon, we
drop this sign notation for the sake of brevity. Separating
explicitly the $\Lb$-particle--proton-hole states, we have

$ $

\be\label{lpweakvertex}
\setlength{\unitlength}{1mm}
\parbox{25mm}{\begin{fmfgraph*}(25,20)
\fmfleftn{l}{2}
\fmfrightn{r}{2}
\fmf{scalar}{l1,ol,l2}
\fmfpoly{shade}{old,olu,ol}
\fmf{heavy}{olu,r2}
\fmf{fermion}{r1,old}
\fmflabel{$\bar\nu_l$}{l1}
\fmflabel{$l^+$}{l2}
\end{fmfgraph*}}
\,\,\,=\,\,\,
\setlength{\unitlength}{1mm}
\parbox{25mm}{\begin{fmfgraph*}(25,20)
\fmfleftn{l}{2}
\fmfrightn{r}{2}
\fmf{scalar}{l1,ol,l2}
\fmfpoly{empty}{old,olu,ol}
\fmf{heavy}{olu,r2}
\fmf{fermion}{r1,old}
\fmflabel{$\bar\nu_l$}{l1}
\fmflabel{$l^+$}{l2}
\end{fmfgraph*}}
\,\,\,+\,\,\,
\setlength{\unitlength}{1mm}
\parbox{40mm}{\begin{fmfgraph*}(40,20)
\fmfleftn{l}{2}
\fmfrightn{r}{2}
\fmf{scalar}{l1,ol,l2}
\fmfforce{(0.2w,0.5h)}{ol}
\fmfforce{(0.8w,0.5h)}{or}
\fmfpoly{shade}{old,olu,ol}
\fmfpoly{shade}{or1,or2,oru,ord}
\fmfforce{(0.35w,0.65h)}{olu}
\fmfforce{(0.35w,0.35h)}{old}
\fmf{heavy,left=.5,tensio=.5}{olu,oru}
\fmf{fermion,left=.5,tension=.5}{ord,old}
\fmf{heavy}{or2,r2}
\fmf{fermion}{r1,or1}
\fmflabel{$\bar\nu_l$}{l1}
\fmflabel{$l^+$}{l2}
\end{fmfgraph*}}
\,\,\, ,\ee

$ $

\noindent
where

$ $

\be\label{barelpweak}
\setlength{\unitlength}{1mm}
\parbox{25mm}{\begin{fmfgraph*}(25,20)
\fmfleftn{l}{2}
\fmfrightn{r}{2}
\fmf{scalar}{l1,ol,l2}
\fmfpoly{empty}{old,olu,ol}
\fmf{heavy}{olu,r2}
\fmf{fermion}{r1,old}
\fmflabel{$\bar\nu_l$}{l1}
\fmflabel{$l^+$}{l2}
\end{fmfgraph*}}
\,\,\, = \,\,\,
\parbox{25mm}{\begin{fmfgraph*}(25,20)
\fmfleftn{l}{2}
\fmfrightn{r}{2}
\fmf{scalar}{l1,ol,l2}
\fmf{fermion}{r1,or}
\fmf{heavy}{or,r2}
\fmf{dbl_dots}{ol,or}
\fmfv{d.sh=di,d.fi=full,d.si=2thick}{ol}
\fmfv{d.sh=c,d.fi=full,d.si=2thick}{or}
\fmflabel{$\bar\nu_l$}{l1}
\fmflabel{$l^+$}{l2}
\end{fmfgraph*}}
\,\,\,+\,\,\,
\parbox{25mm}{\begin{fmfgraph*}(25,20)
\fmfleftn{l}{2}
\fmfrightn{r}{2}
\fmf{scalar}{l1,o,l2}
\fmf{fermion}{r1,o}
\fmf{heavy}{o,r2}
\fmfv{d.sh=sq,d.fi=full,d.si=2thick}{o}
\fmflabel{$\bar\nu_l$}{l1}
\fmflabel{$l^+$}{l2}
\end{fmfgraph*}}
\,\,\, .\ee

$ $

\noindent
The shaded block in Eq.~(\ref{lpweakvertex}) is the full $\Lb p$
interaction amplitude in cold nuclear matter, which is obtained via
dressing a bare $\Lb p$ interaction by $\Lb$-particles--proton-hole
loops

$ $

\be\label{lbpinttot}
\setlength{\unitlength}{1mm}
\parbox{25mm}{\begin{fmfgraph*}(25,20)
\fmfleftn{l}{2}
\fmfrightn{r}{2}
\fmfpolyn{shade}{P}{4}
\fmf{fermion}{r1,P1}
\fmf{heavy}{P2,r2}
\fmf{heavy}{l2,P3}
\fmf{fermion}{P4,l1}
%\fmfv{l=$j$,l.a=120,l.d=3thick}{P4}
\end{fmfgraph*}}
\,\,\,=\,\,\,
\setlength{\unitlength}{1mm}
\parbox{25mm}{\begin{fmfgraph*}(25,20)
\fmfleftn{l}{2}
\fmfrightn{r}{2}
\fmfpolyn{empty,pull=1.4,smooth}{P}{4}
\fmf{fermion}{r1,P1}
\fmf{heavy}{P2,r2}
\fmf{heavy}{l2,P3}
\fmf{fermion}{P4,l1}
\end{fmfgraph*}}
\,\,\,+\,\,\,
\setlength{\unitlength}{1mm}
\parbox{50mm}{\begin{fmfgraph*}(50,20)
\fmfleftn{l}{2}
\fmfrightn{r}{2}
\fmfpolyn{empty,pull=1.4,smooth}{P}{4}
\fmfpolyn{shade}{Pr}{4}
%% legs
\fmf{heavy}{l2,P3}
\fmf{fermion}{P4,l1}
\fmf{fermion}{r1,Pr1}
\fmf{heavy}{Pr2,r2}
%%% internal
\fmf{fermion,left=.5,tension=.5}{Pr4,P1}
\fmf{heavy,left=.5,tension=.5}{P2,Pr3}
\end{fmfgraph*}}
\,\,\,,\ee

$ $

\noindent
where the bare $\Lb$--proton-hole interaction is presented as

$ $

\be\label{lbpbare}
\setlength{\unitlength}{1mm}
\parbox{25mm}{\begin{fmfgraph*}(25,20)
\fmfleftn{l}{2}
\fmfrightn{r}{2}
\fmfpolyn{empty,pull=1.4,smooth}{P}{4}
\fmf{fermion}{r1,P1}
\fmf{heavy}{P2,r2}
\fmf{heavy}{l2,P3}
\fmf{fermion}{P4,l1}
\end{fmfgraph*}}
\,\,\,=\,\,\,
\parbox{25mm}{\begin{fmfgraph*}(25,20)
\fmfleftn{l}{2}
\fmfrightn{r}{2}
\fmf{heavy}{l2,ol}
\fmf{fermion}{ol,l1}
\fmf{fermion}{r1,or}
\fmf{heavy}{or,r2}
\fmf{dbl_dots}{ol,or}
\fmfv{d.sh=c,d.filled=full,d.si=2thick}{ol}
\fmfv{d.sh=c,d.filled=full,d.si=2thick}{or}
\end{fmfgraph*}}
\,\,\,+\,\,\,
\parbox{25mm}{\begin{fmfgraph*}(25,20)
\fmfleftn{l}{2}
\fmfrightn{r}{2}
\fmf{heavy}{l2,o}
\fmf{fermion}{o,l1}
\fmf{fermion}{r1,o}
\fmf{heavy}{o,r2}
\fmfv{d.sh=sq,d.fi=shaded,d.si=4thick}{o}
\end{fmfgraph*}}
\,\,\,.\ee

$ $

The dotted line is determined by Eq.~(\ref{ksdressed}) and the
shaded box represents the short-range $\Lb$--proton-hole
interaction, given in Eq.~(\ref{tln}).

Calculation of diagrams (\ref{opts}) according to standard
diagramatic rules results in the sum of Eqs~(\ref{nucros}) and
(\ref{nucroslb}). Thus, making use of the optical theorem allows to
naturally avoid the double counting problem.

\subsection{Non-strange channel}
There are other processes with the positron or positive muon
production in neutrino nucleus scattering, supplemented by
production of non-strange particles. These give the background to
the above considered processes. This background can be in principle
subtracted by simultaneous  registration of  strange particles in
the final state. However, even without this experimentally
complicated approach, one can hope to detect contributions of
processes with strangeness production.

Let us consider the non-strange processes in more detail. Some of
them are easily distinguishable from those with strangeness
production having different kinematics but the process $\bar{\nu}_l
\rightarrow l^+ +\pi^{-}$ considered in \cite{sawyerpion} and the
related processes $\bar{\nu}_{l} +p\rightarrow l^+ +n$,
$\bar{\nu}_{l} +N\rightarrow l^+ +\Delta (1232)$, considered in
\cite{kim96,kim94}, occur in the same energy--momentum region.
Their probability is rather large.

In terms of the optical theorem formalism these processes can be
interpreted as an excitation of in-medium particle--hole and
$\Delta$--hole quanta of interaction. As well known, cf.
Ref.~\cite{migdal}, these quanta are strongly mixed with each other
and with in-medium pionic excitations. Thus, in medium these
processes should be considered within the optical theorem formalism
to prevent a possible  double counting. Calculating the rates of
the non-strange processes above one has to take into account the
weak-interaction renormalization due to the short-range $NN$ and
$N\Delta$ correlations.

In Ref.~\cite{kim96,kim94}  these effects were partially included
for the $\bar{\nu}_{l} +p\rightarrow l^+ +n$ channel in the
framework of the RPA approximation. The net effect from the
correlations is a suppression by the factor $\simeq0.5$. In the
$\bar{\nu}_{l} +N\rightarrow l^+ +\Delta (1232)$ channel no
correlations were included.

We expect that a more consistent account for correlations will lead
to a larger suppression effect. Indeed, a rough estimation beyond
RPA gives for the reaction $\bar{\nu}_{l} +p\rightarrow l^+ +n$ the
correlation factor, e.g., in the axial current vertex
$\gamma_N(g')\simeq 1/\bigl( 1+ g'\,2\,
\mn^*\,p_{\rm F}(\ro)/\pi^2\bigr)$ at small transverse frequencies
and transverse momenta $\sim p_{\rm F}(\ro)$, being the Fermi
momentum of a nucleon at the normal nuclear density. Evaluating
this expression with $g'\approx 0.7\, m_\pi^{-2}$, being the
spin-isospin Landau-Migdal parameter of the short-range $NN$
interaction, we find 
the suppression factor for the reaction rate about
$\gamma_N^2(g')\simeq$ 0.1--0.3\,. For the reaction $\bar{\nu}_{l}
+N\rightarrow l^+ +\Delta (1232)$, effects due to the $NN$ and
$N\Delta$ correlations are less important, and the resulting
suppression factor is expected to be in the range 0.5--0.7.

To compare the rates of strange and non-strange channels of
anti-neutrino nucleus scattering we take the results from
Ref.~\cite{kim96}, the thin dashed curve in Fig.~3, which are close
to our calculation with account for the baryon mass reduction due
to the mean field interactions.  With the suppression factors above
we estimate that the strange and non-strange channels give
contributions of the same order to the angular integrated cross
section. The cross sections taken at the fixed $\bar\nu$--$l^+$
scattering angle correspond to the different kinematics and can be
distinguished, thereby.

\section{Conclusion}
We calculated the differential cross section for the anti-neutrino
induced production of positive leptons on a nucleus associated with
production of strangeness $S=-1$.

The most important contribution is found to be given by the
reaction $\bar\nu+p\to \Lb + l^+$, which we calculated, including
weak-interaction renormalization in nuclear matter due to the
short-range $\Lb p$ correlations taken within Landau--Migdal
parameterization. The in-medium effects alter essentially the
differential cross section at small $\bar\nu$--$l^+$  scattering
angles both in the absolute value and in the shape. The total cross
section changes by $\sim$10\%.

We also considered a contribution from the in-medium kaon
production process $\bar\nu_l\to l^++K^-$. For that we evaluated
the $K^-$ spectral density  in nuclear matter and included the
weak-coupling vertex renormalization. The latter increases the rate
of positron production in this channel by factor $\sim$10$^5$
compared to that estimated with the free weak coupling. In spite of
that  the contribution  of the reaction channel $\bar\nu_l\to
l^++K^-$ to the full $S=-1$ strange particle rate is $\sim 10^3$
times smaller than that of the reaction $\bar\nu_l+p\to l^+ +\Lb$.
Thus, only a peculiar experimental analysis could allow to
discriminate the contribution from this in-medium $K^-$ channel.

We demonstrated that the rate of both reactions $\bar\nu_l\to
l^++K^-$ and $\bar\nu_l+p\to l^+ +\Lb$ is not given by the direct
sum of the squared matrix elements of the corresponding Feynman
diagrams. Otherwise, some processes would be counted twice. We show
that this double counting problem is easily avoided in the
framework of the optical theorem formalism \cite{vossen,kvaop}. The
closed diagram method, we demonstrated, is quite general  and can
be applied for any other reaction channels, e.g., for the
non-strange reaction channel, which also requires consistent
inclusion of in-medium effects.

Both strange and non-strange contributions to the angular
integrated cross sections are found to be of the same order of
magnitude. However, they are related to the distinct kinematic
regions at the fixed neutrino lepton scattering angle. They also
can be distinguished  with the help of a simultaneous
identification of  strange particles in the final state.

The developed formalism can be utilized in investigation of  other
weak processes $e^-\to K^-+\nu_e$, $e^-+n\to
\Sigma^- +\nu_e $, and $n+n\to \Lb + n$
important for neutron star physics, giving rise to hyperonization
\cite{hyper} and $K^-$ condensation \cite{kcond,kvknucl}. The
corresponding neutrino radiation can result in some observable
consequences, as a jump in neutrino radiation and reheating. The
rates of these processes are sensitive to in-medium renormalization
of the  weak-interaction vertices, $\Lb$--nucleon correlation
effects, and to the $K^-$ spectral density as well.

\vspace*{1cm}

{\bf{Acknowledgments.}} The authors would like to thank R. Dahl, M. Lutz and
W. Weinhold for discussions and helpful remarks, and the GSI theory
group for hospitality and support. The work has been supported in
part by BMBF under the program on scientific-technological
collaboration (WTZ project RUS-656-96).

%%%%%%%%%%%%%%%%%%%%%%%%%%%%%%%%%%%%%%

\newpage
%%%%%%%%%%%%%%%%%%%%%%% FIG. 1 %%%%%%%%%%%%%%%%%%%%
%%%%%%%%%%%%%%%%%%%%% FIG. 2 %%%%%%%%%%%%%%%%%%%%%%%%
%%%%%%%%%%%%%%%%%%%%%%% FIG. 3 %%%%%%%%%%%%%%%%%%%%%%%
%%%%%%%%%%%%%%%%%%%%%%%%%%%%%%%%%%%%%%%%%%%%%%%%%%%%%%%
\end{fmffile}
\end{document}